\documentclass[aps,superscriptaddress,twocolumn,a4paper,final,1p,times]{revtex4-1}
\usepackage[dvipdfmx]{graphicx}
\usepackage{amsmath,amssymb}
\usepackage{color}
\usepackage{hyperref}
\usepackage{float}
\usepackage{wrapfig}
\usepackage{textcase}
\newcommand{\Slash}[1]{{\ooalign{\hfil/\hfil\crcr$#1$}}}
\allowdisplaybreaks[1]
\begin{document}
\title{In medium $\eta^\prime$ mass and $\eta^\prime N$ interaction \\ based on chiral effective theory}
\author{Shuntaro Sakai}
\affiliation{Department of Physics, Graduate School of Science, Kyoto University, Kyoto 606-8502, Japan}
\author{Daisuke Jido}
\affiliation{Department of Physics Tokyo Metropolitan University Hachioji, Tokyo 192-0397, Japan}
\begin{abstract}
The in-medium $\eta^\prime$ mass and the $\eta^\prime N$ interaction are investigated in an effective theory based on the linear realization of the SU(3) chiral symmetry.
We find that a large part of the $\eta^\prime$ mass is generated by the spontaneous breaking of chiral symmetry through the U$_A$(1) anomaly.
As a consequence of this observation, the $\eta^\prime$ mass is reduced in nuclear matter where chiral symmetry is partially restored.
In our model, the mass reduction is found to be 80MeV at the saturation density.
Estimating the $\eta^\prime N$ interaction based on the same effective theory, we find that the $\eta^\prime N$ interaction in the scalar channel is attractive sufficiently to form a bound state in the $\eta^\prime N$ system with a several MeV binding energy.
We discuss the origin of attraction by emphasizing the special role of the $\sigma$ meson in the linear sigma model for the mass generation of $\eta^\prime$ and $N$.
\end{abstract}
\maketitle

\section{\label{intro}Introduction}
The $\eta^\prime$ meson has a large mass compared to the other
pseudoscalar mesons, like $\pi, K, \eta$.
The mass spectrum of the low lying pseudoscalar mesons has been
discussed as the U$_A$(1) problem \cite{Weinberg1975}.
The mass of $\eta^\prime$ can be explained by the U$_A$(1)
anomaly in QCD \cite{Witten1979,Veneziano1979}.
The quantum anomaly is the phenomenon that symmetries in the classical
level are broken by quantum effect.
The QCD lagrangian is invariant under the U$_A$(1) transformation for
the quark field, but the symmetry is broken explicitly by the quark loop
effect and the divergence of the U$_A$(1) current does not vanish \cite{Bardeen1969}.
When chiral symmetry is broken spontaneously, the non-zero
divergence of the U$_A$(1) current permits the non-vanishing mass of the
pseudoscalar flavor-singlet meson even in the chiral limit.

The medium effect to the $\eta^\prime$ mass through the effective U$_A$(1) restoration has been discussed.
The effective U$_A$(1) restoration is caused by the in-medium decrease
of the instanton density \cite{Gross1981,Kapsta1996}.
The reduction of the instanton density in the medium may lead to the
suppression of the expectation value of the U$_A$(1) current divergence in
the medium.
The vanishing expectation value
of the U$_A$(1) current for the vacuum and $\eta^\prime$ states
forces to make the $\eta^\prime$ meson to be massless in the same way as
the other pseudoscalar mesons.

Apart from the effective U$_A$(1) restoration, as we will discuss later in detail, the chiral symmetry
breaking is indispensable to the mass difference between the
pseudoscalar flavor-singlet and flavor-octet mesons in addition to the
U$_A$(1) anomaly.
Recently, the reduction of the absolute value of the quark condensate,
which is called as partial restoration of chiral symmetry, in the
nuclear medium has been discussed intensively from the theoretical and
experimental points of view and it is suggested by the analysis of the
experimental data of pionic atoms that the partial restoration does take
place in nuclei actually \cite{Suzuki2004}.
If one takes account of the necessity of the chiral symmetry breaking in
the generation of the $\eta^\prime$ mass, it is expected that the
flavor-singlet meson mass decreases in the nuclear medium, in which
chiral symmetry is partially restored \cite{Jido2012}.

There are many theoretical works \cite{Kapsta1996,
Pisarski1984, Bernard1987, Hatsuda1994, Saito2007, Costa2003, Bass2006,
Nagahiro2005, Nagahiro2012,Nagahiro2006,Oset2011,Benic2011,Kwon2012,Lee2013} and experimental attempts
\cite{Moskal2000a, Moskal2000b, Nanova2012, Csorgo2010} involved in
the in-medium $\eta^\prime$ properties from the various points of view.
Particularly, the effect of the chiral symmetry to the $\eta^\prime$ meson is
discussed in \cite{Jido2012,Cohen1996,Lee1996}.

The mass reduction of $\eta^\prime$ in the nuclear medium implies that
the $\eta^\prime$ meson feels attraction in the nuclear medium because
the mass modification is represented by the self energy of the meson in
the medium and the self energy turns out to be the optical potential in
the non-relativistic limit.
The attraction in the nuclear matter suggests an attractive $\eta^\prime N$
two-body force as an elementary interaction.
If it is enough strong, we expect a $\eta^\prime N$ bound state.
This is an analogous state of $\Lambda$(1405), which is
considered as a bound state of $\bar{K}N$.

So far, the interaction between $\eta^\prime$ and $N$ is not known.
We do not know even whether it is attractive or repulsive.
There are some experimental suggestions about the $\eta^\prime N$
scattering length and the in-medium $\eta^\prime$ properties.
From the $pp\rightarrow pp\eta^\prime$ process, the scattering length of
$\eta^\prime p$ has been extracted and its value has been estimated about 0.8fm
\cite{Moskal2000a} or 0.1fm with the sign undetermined \cite{Moskal2000b}.
The absorption of $\eta^\prime$ into nuclei has been extracted by the $\gamma
p\rightarrow \eta^\prime p$ process in nuclei and the absorption of
$\eta^\prime$ is relatively small compared to that of the $\omega$ meson
\cite{Nanova2012}.
These experimental data  suggest the weakness of the
$\eta^\prime N$ interaction.
On the other hand, large mass reduction of $\eta^\prime$ has been reported
from the analysis of the low-energy pion distribution in the
relativistic heavy ion collision \cite{Csorgo2010}.
This suggests the strong attraction between $\eta^\prime N$ if one
considers that this mass reduction occurs due to the partial restoration of
chiral symmetry.
The $\eta^\prime N$ interaction and in-medium properties of
$\eta^\prime$ should be understood in a unified manner and the
theoretical study concerning the $\eta^\prime N$ interaction is progressing
\cite{Oset2011}.

In this paper, taking partial restoration of chiral symmetry in the
nuclear medium as a basis of our argument, we estimate the amount of the
expected $\eta^\prime$ mass reduction in the nuclear medium and the two body
interaction strength of $\eta^\prime N$ in vacuum.
A preliminary account of this work has been reported in a paper of the
conference proceedings \cite{Jido2012a}.
In this paper, we explain fully the details of the model which we use
and the calculation method.
We also discuss the dependence of the results on the model.
In Sec.\ref{massred1}, we explain the relation of the $\eta^\prime$ meson and
the chiral symmetry breaking.
In Sec.\ref{lsm}, we introduce an effective lagrangian for the
$\eta^\prime$ meson in the nuclear medium based on the linear sigma
model, and evaluate the in-medium mass reduction of the $\eta^\prime$.
In Sec.\ref{epnbs1}, we show the obtained $\eta^\prime
N$ interaction strength and the binding energy and the scattering length
of the $\eta^\prime N$ in vacuum.
The conclusion and some remarks of this paper is given in Sec.\ref{conc}.

\section{\label{massred1}The relation between \texorpdfstring{$\eta^\prime$}{eta'} meson and chiral symmetry}
The mass difference between the $\eta$ and $\eta^\prime$ mesons has been discussed based on the QCD partition function
\cite{Cohen1996,Lee1996} or the SU(3) chiral symmetry \cite{Jido2012}.
The U$_A$(1) symmetry is broken explicitly due to the quantum effect.
Therefore, with the spontaneous chiral symmetry breaking, the
$\eta^\prime$ meson can have a finite mass even in the chiral limit
contrary to the other pseudoscalar NG bosons.
But, the U$_A$(1) anomaly effect lifting the $\eta^\prime$ meson
mass in vacuum cannot affect the pseudoscalar mass
spectrum when chiral symmetry is restored.
This is because the $\eta$ and $\eta^\prime$ mesons masses should degenerate
in the SU(3) chiral symmetric phase even if U$_A$(1) symmetry is
explicitly broken by the anomaly effect according to
Refs. \cite{Cohen1996,Lee1996,Jido2012}.

In the following, we explain the mechanism of the degeneracy of the pseudoscalar
flavor singlet and octet mesons based on the SU(3) chiral symmetry
\cite{Jido2012}.
We consider the 3 flavor chiral symmetry SU(3)$_L\otimes$SU(3)$_R$, and
we assume that the effect of the change of the instanton density near
the normal nuclear density to the $\eta^\prime$ mass is small compared
to the effect of partial restoration of chiral symmetry.

First, we define the transformation properties of the quark field under the
SU(3)$_L \otimes$SU(3)$_R$ transformation.
The left-handed quark $q_L$ and the right-handed quark $q_R$ are defined as
\begin{eqnarray}
 q_L&=&\frac{1-\gamma_5}{2}q,\\
 q_R&=&\frac{1+\gamma_5}{2}q.
\end{eqnarray}
Because the quark fields, $q_L$ and $q_R$, belong to the fundamental
representation of SU(3)$_L$ and SU(3)$_R$ respectively, the
transformation properties of the quark fields under SU(3)$_L \otimes$SU(3)$_R$ is
written as
\begin{equation}
q_i\rightarrow e^{i\theta_i^a \lambda^a/2}q_i\ \ \ (i=L,R).
\end{equation}
Here, $\lambda^a$ ($a=1\sim8$) is the Gell-Mann matrix.

The QCD lagrangian is invariant under the SU(3)$_L \otimes$SU(3)$_R$
transformation in the chiral limit.
When $\theta_R=\theta_L\equiv \theta_V$, the transformation for the quark
field $q$ is written as 
\begin{eqnarray}
 q \rightarrow e^{i\theta^a_V \lambda^a/2}q.\label{qvtrans}
\end{eqnarray}
We call this transformation as vector transformation.
When $\theta_R=-\theta_L \equiv \theta_A$, the transformation for the
quark field $q$ is written as
\begin{eqnarray}
 q \rightarrow e^{i\theta^a_A \gamma_5\lambda^a/2}q.
\end{eqnarray}
We call this transformation as axial transformation.
For an infinitesimal transformation, the quark field transforms as
\begin{equation}
 q\rightarrow \left(1+i\theta_A^a \gamma_5\frac{\lambda^a}{2}\right)q.\label{eisat}
\end{equation}
This implies 
\begin{equation}
 \left[Q_A^a,q\right]=-\frac{1}{2}\lambda^a\gamma_5 q \label{qaxialtrans}
\end{equation}
and
\begin{eqnarray}
\left[ Q_A^a,\bar{q} \right]=-\frac{1}{2}\bar{q}\lambda^a \gamma_5
\end{eqnarray}
with the generator of the axial transformation $Q_A^a$.

Under the SU(3)$_L$ $\otimes$ SU(3)$_R$ symmetry, the hadron fields
can be classified in terms of the irreducible representation of
SU(3)$_L\otimes$SU(3)$_R$.
Assuming that the mesons are composed of the quark bilinear form and
parity invariance is satisfied in vacuum, the meson fields belong to the
$\left({\bf 3}_L, \bar{{\bf 3}}_R\right)\oplus \left(\bar{\bf 3}_L, {\bf
3}_R\right)$ representation.
In terms of the vector transformation, the meson fields belonging to $({\bf 3}_L,\bar{\bf 3}_R)\oplus(\bar{\bf 3}_L,{\bf 3}_R)$ can be decomposed into
the octet and singlet representations being the irreducible
representation of SU(3)$_V$, with the fact of ${\bf 3}\otimes\bar{\bf 3}={\bf
8}\oplus{\bf 1}$.
Taking the meson fields as the parity eigenstates, one can obtain the
parity even mesons as
$\frac{1}{\sqrt{2}}(\bar{q}_L\frac{1}{\sqrt{3}}q_R+\bar{q}_R\frac{1}{\sqrt{3}}q_L)=\frac{1}{\sqrt{6}}\bar{q}q$,
$\frac{1}{\sqrt{2}}(\bar{q}_L\frac{\lambda_a}{2}q_R+\bar{q}_R\frac{\lambda_a}{2}q_L)=\frac{1}{\sqrt{2}}\bar{q}\frac{\lambda_a}{2}q$ and the
parity odd mesons as
$\frac{i}{\sqrt{6}}(\bar{q}_Lq_R-\bar{q}_Rq_L)=\frac{1}{\sqrt{6}}\bar{q}i\gamma_5q$,
$\frac{i}{\sqrt{2}}(\bar{q}_L\frac{\lambda_a}{2}q_R-\bar{q}_R\frac{\lambda_a}{2}q_L)=\frac{1}{\sqrt{2}}\bar{q}\frac{\lambda_a}{2}i\gamma_5q$.
We assign the pseudoscalar octet mesons ($\pi,\ K,\ \eta_8$) to
\textbf{8} and singlet ($\eta_0$) to \textbf{1}, so the 9 pseudoscalar
mesons are settled into a part of the same representation of SU(3)$_L$
$\otimes$SU(3)$_R$.
In the real world, the $\eta$ and $\eta^\prime$ mesons are mixed
states of $\eta_0$ and $\eta_8$ owing to the flavor SU(3)
symmetry breaking, and their masses are obtained by diagonalizing their
mass matrix.
As the same way, the scalar mesons ($\sigma_0,a_0,\kappa,\sigma_8$) are
assigned to the rest part of the $({\bf 3}_L,{\bf 3}_R)\otimes ({\bf 3}_R,{\bf 3}_L)$ representation of SU(3)$_L$ $\otimes$SU(3)$_R$.
From these assignment, the 18 scalar and pseudoscalar mesons belong to
the same chiral multiplet of SU(3)$_L \otimes$SU(3)$_R$.

If one considers the SU(3)$_L \otimes$SU(3)$_R$ transformation, the $\eta_0$
meson can be transformed to other pseudoscalar mesons like $\pi$ or $K$,
$\eta_8$.
The singlet and octet are irreducible representations in SU(3)$_V$, so
the vector transformation alone cannot transform the singlet $\eta_0$ into the
pseudoscalar octet mesons.
In contrast, the axial transformation can mix the singlet and
octet mesons because the axial transformation is not the element of
SU(3)$_V$ but that of SU(3)$_L \otimes$SU(3)$_R$.
Thus, the decomposition into singlet and octet makes sense
when chiral symmetry is broken, while they are transformed each other with axial
transformations in the case that chiral symmetry exists.

Here, we demonstrate the transformation between these 9 pseudoscalar mesons with
the SU(3)$_L \times$ SU(3)$_R$ transformation explicitly.
Using Eq.(\ref{qaxialtrans}), the flavor singlet pseudoscalar meson field
$\eta_0=\bar{q}i\frac{\gamma_5}{\sqrt{6}}q$ is transformed as
\begin{eqnarray}
 \delta^a\left(\bar{q}i\frac{\gamma_5}{\sqrt{6}}
  q\right)&=&\left[Q_A^a,\bar{q}i\frac{\gamma_5}{\sqrt{6}} q\right]=\bar{q}\left\{\frac{i\gamma_5}{\sqrt{6}},-\frac{\lambda^a}{2}\gamma_5\right\}q\nonumber\\
&=&-\bar{q}i\frac{\lambda^a}{\sqrt{6}} q,\label{atrans1}
\end{eqnarray}
and the obtained octet-scalar meson field is transformed as
\begin{eqnarray}
 \delta^b \left(-\bar{q}i\frac{\lambda^a}{\sqrt{6}}
  q\right)&=&\left[Q_A^b,-\bar{q}i\frac{\lambda^a}{\sqrt{6}}q\right]=\bar{q}\left\{-i\frac{\lambda^a}{\sqrt{6}},-\frac{\lambda^b}{2}\gamma_5\right\}q\nonumber\\
&=&d^{abc}\bar{q}i\gamma_5\frac{\lambda^c}{\sqrt{6}}q.\label{atrans2}
\end{eqnarray}
The Eq.(\ref{atrans1}) shows that the singlet pseudoscalar meson
is transformed into a scalar octet meson through the first axial
transformation and the second axial transformation changes the
flavor-octet scalar meson into a pseudoscalar octet meson in
Eq.(\ref{atrans2}).
Thus, the pseudoscalar flavor singlet and octet mesons are transformed into each other under
the SU(3)$_L \otimes$SU(3)$_R$ transformations.
This means that the $\eta_0$ meson degenerates to the other pseudoscalar mesons
when chiral symmetry exists.

Here, it is notable that the degeneracy of
the singlet and the octet mesons do not necessarily happen in the $N_f=2$ case.
The case of $N_f=2$ corresponds to the limit that the strange quark
mass, $m_s$, goes to the infinity in $N_f=3$.
So the SU(3)$_L \otimes$SU(3)$_R$ symmetry is strongly broken.
Hence, the mass degeneracy of the $\eta^\prime$ and pseudoscalar octet
mesons does not necessarily take place.
This is consistent with the argument in \cite{Lee1996}.

With simple assumptions, we can estimate the amount of the mass reduction of
$\eta^\prime$ in a nuclear medium.
Here, we take the chiral limit, so $\eta^\prime$ and $\eta$ correspond
to $\eta_0$ and $\eta_8$ respectively.

First, we assume the linear dependence of the mass difference of the
$\eta$ and $\eta^\prime$ on the flavor singlet combination of chiral
condensate.
With this assumption, the mass difference of $\eta$ and $\eta^\prime$ is
written using a constant $C$ as
\begin{equation}
 m_{\eta^\prime}^2-m_{\eta}^2=C\left(2\left<\bar{q}q\right>+\left<\bar{s}s\right>\right)
  \label{ceta}.
\end{equation}
Here, we have taken
$\left<\bar{q}q\right>=\left<\bar{u}u\right>=\left<\bar{d}d\right>$.
From Eq.(\ref{ceta}), $C$ can be written as $C=\frac{m_{\eta^\prime}^2-m_\eta^2}{2\left<\bar{q}q\right>+\left<\bar{s}s\right>}$.
We suppose that the strangeness condensate $\left<\bar{s}s\right>$ and
the $\eta$ mass do not change so much in the nuclear matter.
Substituting the explicit form of $C$, we obtain
\begin{equation}
 m_{\eta^\prime}^2-m_{\eta^\prime}^{\ast2}=\frac{2}{3}\left(m_{\eta^\prime}^2-m_\eta^2\right)\left(1-\frac{\left<\bar{q}q\right>^\ast}{\left<\bar{q}q\right>}\right),
\end{equation}
where $m_{\eta^\prime}^\ast$ and $\left< \bar{q}q \right>^\ast$ denote the in-medium values of the $\eta^\prime$ mass and the quark condensate, respectively.
With the low density theorem \cite{Drukarev1991}, the reduction of the
quark condensate at the leading order of the density is
written as
\begin{equation}
 \frac{\left<\bar{q}q\right>^\ast}{\left<\bar{q}q\right>}=1-\frac{\sigma_{\pi
  N}}{m_\pi^2 f_\pi^2}\rho+\mathcal{O}(\rho^{4/3}),
\end{equation}
where $\sigma_{\pi N}$ is the $\pi N$ sigma term.
With $m_{\eta^\prime}^\ast=m_{\eta^\prime}-\Delta
m_{\eta^\prime}$ and neglecting $(\Delta m_{\eta^\prime})^2$, we obtain
the mass reduction of $\eta^\prime$ as
\begin{equation}
 \Delta
  m_{\eta^\prime}=\frac{2}{3}\frac{m_{\eta^\prime}^2-m_{\eta}^2}{2m_{\eta^\prime}}\frac{\sigma_{\pi
  N}}{m_\pi^2 f_\pi^2}\rho. 
\end{equation}
Using the observed values of the masses and the decay constant and the
typical value for $\sigma_{\pi N}$, which reproduces the 35\% reduction
of the quark condensate 
at the normal nuclear density, $\Delta m_{\eta^\prime}$ takes a value
around 80 to 100MeV at the normal nuclear density.
%

\section{\label{lsm}Linear sigma model}
To study $\eta^\prime$ in the nuclear matter and the $\eta^\prime N$
interaction in vacuum, we use the linear sigma model as a chiral
effective theory.
The linear sigma model is based on the global symmetry as QCD and
contains the effects of the finite current quark mass and the U$_A$(1)
anomaly \cite{Gell-Mann1960, Schechter1971, Kawarabayashi1980, Lenaghan2000}.
The advantages of the linear sigma model are as follows; it has the mechanism
of spontaneous chiral symmetry breaking and that the
physical quantities are expressed by the sigma condensate, which is the
order parameter of the spontaneous chiral symmetry breaking in the
linear sigma model.
The sigma condensate is given by minimizing the effective potential
calculated from the lagrangian.
The sigma condensate characterizes the vacuum to be realized as the
ground state.
This means that the linear sigma model is a model which can describe the
response of the physical quantities caused by the change of the vacuum.
In the case of the non-linear sigma model, the physical quantities are
written by the low energy constants, which should be, in principle,
determined again by the information of the vacuum in the nuclear medium.
Therefore, the non-linear sigma model is not suitable for the
present aim to directly connect the $\eta^\prime$ mass with partial
restoration of chiral symmetry.
In addition, since hadron is the fundamental degree of freedom in
the linear sigma model, we can introduce the nucleon fields straightforwardly.
This is a different point from the quark based model, such as the NJL
model, in which we have to build up the nucleon within the model.

\subsection{The lagrangian of the linear sigma model}
The lagrangian of the linear sigma model is constructed to possess the same global symmetry of QCD.
The fundamental degree of freedom is hadron.
The hadron fields can be assigned the irreducible representation of
SU(3)$_L \otimes$SU(3)$_R$.
As the result, the transformation properties of the hadron fields under
the SU(3)$_L \otimes$SU(3)$_R$ transformation are fixed and the
lagrangian is constructed so as to be invariant under the transformation.
Chiral symmetry is spontaneously broken with certain parameter sets,
and then the sigma condensate has a non-zero value.
In the following, we explain the lagrangian of the linear sigma model
which we use to calculate the in-medium $\eta^\prime$ mass and the
$\eta^\prime N$ interaction.

\subsubsection{Meson part}
As mentioned above, 
the meson field $M$ belongs to the $({\bf 3}, \bar{\bf 3})$ irreducible representation, which means
that the meson field transforms as {\bf 3} under the SU(3)$_L$
transformation and $\bar{\bf 3}$ under the SU(3)$_R$ transformation.
Thus, the transformation
rule of the meson field under SU(3)$_L\otimes$SU(3)$_R$ is
\begin{equation}
 M \rightarrow LMR^\dagger, \label{mtrans}
\end{equation}
where $L\in SU(3)_L,\ R\in SU(3)_R$.
Here, the scalar and pseudoscalar meson field $M$ is  written in terms of the physical meson fields as
\begin{equation}
 M=M_s+iM_{ps}=\sum^8_{a=0}\frac{\lambda_a\sigma_a}{\sqrt{2}}+i\sum_{a=0}^8\frac{\lambda_a\pi_a}{\sqrt{2}},
\end{equation}
where $\lambda_a$ ($a=1\sim8$) is the Gell-Mann matrix and
$\lambda_0=\sqrt{\frac{2}{3}}{\bf 1}$ with the unit matrix {\bf 1},
which are normalized as 
\begin{equation}
 {\rm tr}\left(\lambda_a\lambda_b\right)=2\delta_{ab}\ \ \ \ (a,b=0\sim8).
\end{equation}
The explicit form of the pseudoscalar meson field is given as
\begin{eqnarray}
&&M_{ps}=\sum_{a=0}^8\frac{\lambda_a\pi_a}{\sqrt{2}}\nonumber\\
&&\ \ =
	\begin{pmatrix}
		\frac{\pi^0}{\sqrt{2}} + \frac{\eta_8}{\sqrt{6}} +\frac{\eta_0}{\sqrt{3}} & \pi^+ &K^+ \\
		\pi^- & -\frac{\pi^0}{\sqrt{2}} + \frac{\eta_8}{\sqrt{6}} +\frac{\eta_0}{\sqrt{3}} & K^0 \\
		K^- & \bar{K}^0 & -\sqrt{\frac{2}{3}}\eta_8 + \frac{\eta_0}{\sqrt{3}}
	\end{pmatrix}.\nonumber\\
\end{eqnarray}

To include the effect of the finite current quark mass, we give the quark mass $\chi$ the fictitious
transformation rule under the SU(3)$_L\otimes$SU(3)$_R$ transformation to
maintain chiral symmetry.
If one assumes the transformation rule of $\chi$ as
\begin{equation}
 \chi \rightarrow L\chi R^\dagger, \label{chitrans}
\end{equation}
the QCD lagrangian is invariant under the SU(3)$_L\otimes$SU(3)$_R$
transformation.
Here, we take the explicit form of $\chi$ as
\begin{equation}
 \chi=\sqrt{3}\begin{pmatrix}
       m_u & & \\
        & m_d & \\
        & & m_s
      \end{pmatrix}
=\sqrt{3}
\begin{pmatrix}
       m_q & & \\
        & m_q & \\
        & & m_s
      \end{pmatrix},
\end{equation}
where $m_u,m_d,m_s$ are the up, down, strange quark masses, respectively.
Taking $m_u=m_d\equiv m_q$, we introduce the isospin symmetry, and we break
the SU(3) flavor symmetry breaking with $m_q \neq m_s$.
Owing to the flavor symmetry breaking, $\left<\sigma_8\right>$ has a non-zero
value as well as $\left< \sigma_0\right>$.

The lagrangian constructed so as to have the same global symmetry as
that of QCD is
\begin{eqnarray}
\mathcal{L}_{\rm meson}&=& \frac{1}{2}\mathrm{tr}(\partial_\mu
 M\partial^\mu M^\dagger)
 -\frac{{\mu}^2}{2}\mathrm{tr}(MM^{\dagger})\nonumber\\
&&-\frac{\lambda}{4}\mathrm{tr}[(MM^{\dagger})^2] -\frac{{\lambda}^{\prime}}{4}[\mathrm{tr}(MM^{\dagger})]^2 \nonumber\\
& & +A\mathrm{tr}\left( {\chi} M^{\dagger} + \chi^\dagger M \right) +\sqrt{3}B\left( \det M + \det M^\dagger \right).\nonumber\\  \label{mesonl}
\end{eqnarray}
This lagrangian has five parameters, $\mu^2,\ \lambda,\ \lambda^\prime,\
A,\ B$, which cannot be fixed only by the symmetry.
We determine them to reproduce the physical quantities.
In this lagrangian, the fifth term with $\chi$ represents the
current quark mass contribution ( or the flavor symmetry breaking ) as
mentioned above.
The last term proportional to $B$ represents the effect of the U$_A$(1) anomaly.
This term corresponds to the Kobayashi-Maskawa-'t Hooft
term \cite{Kobayashi1970,'tHooft1976}.

When chiral symmetry is broken spontaneously, the sigma
condensates, $\left<\sigma_0\right>$ and $\left<\sigma_8\right>$, are non-zero.
In the chiral symmetry broken phase, the meson masses are
written in terms of the sigma condensate because the meson masses
are defined as the curvature mass at the vacuum where the sigma condensate
is non-zero value.
The explicit form of the meson masses in the $\rho=0$ vacuum at tree
level are shown in Appendix \ref{appa}.

We obtain the relation between the sigma condensate and the meson
decay constants from the axial current and the definition of the meson
decay constants.
The octet axial current $A_a^\mu$ $(a=1 \sim 8)$ is calculated with the
Noether theorem as
\begin{eqnarray}
A^\mu_a&=&\frac{\partial\mathcal{L}}{\partial\left(\partial_\mu
					      M\right)}\delta
M_a\nonumber\\
&=&{\rm tr}\left[\partial^\mu M_{ps}\left\{\lambda_a,M_s\right\}-\partial^\mu M_s\left\{\lambda_a,M_{ps}\right\}\right],
\end{eqnarray}
where $\delta M_a=\frac{i}{2}\left\{\lambda_a,
M\right\}$ is the infinitesimal variation of the meson field under the
axial transformation of SU(3)$_L \otimes$SU(3)$_R$.
The definition of the meson decay constant is
\begin{equation}
 \left<0\left| A_\mu^a \right| \pi^b(p) \right>=-ip_\mu
  f_a\delta^{ab}e^{-ip\cdot x}.\label{acurrent}
\end{equation}
Thus, calculating the matrix element of the axial current with
Eq.(\ref{acurrent}), we obtain the relation between the
sigma condensates and the meson decay constants as
\begin{eqnarray}
 f_\pi&=&\sqrt{\frac{2}{3}}\left< \sigma_0\right>+\frac{1}{\sqrt{3}}\left<
			  \sigma_8\right> \label{fpi}\\
f_K&=&\sqrt{\frac{2}{3}}\left< \sigma_0 \right>-\frac{\left< \sigma_8 \right>}{2\sqrt{3}}.\label{fk}
\end{eqnarray}

We discuss the relation of the order parameter of the spontaneous chiral
symmetry breaking in the linear sigma model and QCD.
The quark and hadron quantities can be related by the ansatz that the
symmetry property should be shared by both QCD and the linear
sigma model.
In the linear sigma model parameter, $\chi$ represents the quark mass.
Equating the derivatives of the partition functions of QCD and linear
sigma model with respect to the quark mass, we obtain the relation
between the quark and sigma condensates at the tree level as
\begin{eqnarray}
 \left<\bar{q}q\right> &=& -2A\left( \left<\sigma_0\right>+\frac{\left<\sigma_8\right>}{\sqrt{2}}\right)\\
\left<\bar{s}s\right> &=& -2A\left( \left<\sigma_0\right>-\sqrt{2}\left<\sigma_8\right>\right).
\end{eqnarray}

The parameters in the lagrangian are determined so as to reproduce the
physical values of the meson masses, the meson decay constants and the
$u,d$ quark mass $m_q$.
The details of parameter fixing are given in the Appendix \ref{appb}.

In this paper, we do not consider the density dependence of the parameters.
The dependence of parameter $B$, which represents the effect of the
U$_A$(1) anomaly, is also responsible for the mass reduction of $\eta^\prime$.
The density dependence of the parameter $B$ is discussed by the
instanton-liquid model and the effect of the anomaly decreases in the
nuclear matter \cite{Shuryak1982}.
Thus, the calculation in this paper gives a lower bound of the
$\eta^\prime$ mass reduction.

\subsubsection{\label{baryonsec}Baryon part}
To consider the change of the meson properties in the nuclear medium, we
introduce the nucleon field to the lagrangian of the meson fields of the
SU(3) linear sigma model.
The transformation property of baryon is not unique even if one regards the
baryon as a composite object of the three quarks.
The baryon representations which are allowed within the symmetry are
(\textbf{3},$\bar{{\bf3}}$)$\oplus$($\bar{\textbf{3}}$,{\bf 3}) and
(\textbf{8},{\bf 1})$\oplus$({\textbf{1},{\bf 8}) \cite{Christos1987}.
Here, we use the (\textbf{3},$\bar{{\bf3}}$)$\oplus$($\bar{\textbf{3}}$,{\bf 3}) representation.
The lagrangian is written as the following;
\begin{eqnarray}
\mathcal{L}_{\rm baryon}&=&\bar{\psi}\left(i\Slash{\partial}-m_N\right)\psi
 -g\bar{\psi}\left(\frac{\tilde{\sigma}_0}{\sqrt{3}}{\bf
	      1}+\frac{\tilde{\sigma}_8}{\sqrt{6}}{\bf 1
	      }\right)\psi\nonumber\\
&&\ \ \ \ \ \ -g\bar{\psi}i\gamma_5\left(\frac{\vec{\pi}\cdot\vec{\tau}}{\sqrt{2}}+\frac{\eta_{0}}{\sqrt{3}}{\bf 1}+\frac{\eta_8}{\sqrt{6}}{\bf
1}\right)\psi,\nonumber\\\label{lagrangian}
\end{eqnarray}
where $\vec{\tau}=\left(\tau_1,\tau_2,\tau_3\right)$, $\tau_i \
(i=1\sim 3)$ are Pauli matrices, $\sigma_i=\left< \sigma_i
\right>+\tilde{\sigma_i}$, {\bf 1} is $2\times2$ unit matrix in the
flavor space and $m_N$ is the nucleon mass.
The nucleon fields are represented as
\begin{eqnarray}
 \psi &=&\begin{pmatrix}
	p \\
	n \\
       \end{pmatrix},
\end{eqnarray}
and the nucleon mass $m_N$ is given by the spontaneous breaking of
chiral symmetry as
\begin{equation}
 m_N=\frac{g}{\sqrt{3}}\left(\left<\sigma_0\right>+\frac{\left<\sigma_8\right>}{\sqrt{2}}\right)\label{mnucl}
\end{equation}
Here, we have showed only the relevant terms for the following calculation here.

The free parameter involved in the lagrangian of the baryon part is the
coupling constant $g$.
This parameter $g$ can be determined from the observation that the
quark condensate reduces about 35\% at the normal density \cite{Suzuki2004}.

In the following, we mention the nucleon mass in the linear sigma model.
The parameter $g$ determined by the magnitude of partial restoration of
chiral symmetry is so small that the
nucleon mass in vacuum can not be reproduced.
On the other hand, if we determine the $g$ so as to reproduce the in-vacuum
nucleon mass, the $g$ is too large to restore chiral symmetry fully at
the densities lower than the saturation density.
This problem is known as Lee-Wick singularity \cite{Lee1974}.
This inconsistency can be solved, for instance, by introducing the parity
doublet baryon \cite{DeTar1989, Jido1998av, Kim1998upa,Sasaki2011},
where a part of the nucleon mass comes from a chiral invariant mass term
rather than the spontaneous breaking of chiral symmetry.
According to the low energy theorem, the interaction between the
pseudoscalar meson and baryon is not dependent on the representation of
the baryon in SU(3)$_L \otimes$SU(3)$_R$ when chiral symmetry is
spontaneously broken.
So, we expect that the following calculations are not affected by how we
introduce the baryon in the linear sigma model as long as we keep chiral
symmetry.

\subsection{The vacuum condition and the medium effect}\label{med1}
\begin{figure}[t]
\includegraphics[width=8cm,clip]{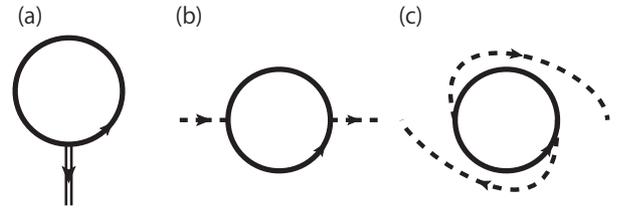}
\caption{\label{1loop}The medium effect of the nucleon one loop to the meson mass.
The solid line, double-solid line and dashed line denote a nucleon,
 scalar meson and pseudoscalar meson, respectively. Diagram (a)
 contributes to the determination of the vacuum. Diagram (b) and (c) are used in the calculation of the in-medium meson mass.}
\end{figure}
In the linear sigma model, the vacuum is defined by the minimum point of
the effective potential.
In this paper, we evaluate the effective potential with the nucleon one loop approximation.
The one loop diagrams considered in this work are given in Fig.\ref{1loop}.
To include the medium effect, we calculate these one loop diagrams using the
nucleon propagator with the Pauli blocking effect.
The nucleon propagator is given as
\begin{eqnarray}
P_{\rm
 med}(p)&=&(\Slash{p}+m_N)\left\{\frac{i}{p^2-m_N^2+i\epsilon}\right.\nonumber\\
&&\left.-2\pi\delta(p^2-m_N^2)\theta(p_0)\theta(k_f-|\vec{p}|)\right\}.
\end{eqnarray}
In the calculation, we regard the nucleon mass as very large
and take the leading term of $1/m_N$.
Diagram (a) of Fig.\ref{1loop} contributes to the
determination of the vacuum and diagram (b) and (b) give the
in-medium self-energy of the meson and the explicit $\rho$ dependence to
the meson mass.
We write the contribution to the effective potential from the first diagram of the
Fig.\ref{1loop} as $V_{MF}(\rho)$ and the contribution to the meson
mass from the second and third diagrams as $\Sigma_{ph}(\rho)$.
Using the propagator including the Pauli blocking effect, $V_{MF}(\rho)$
is calculated as
\begin{equation}
 V_{MF}(\rho)=\frac{g\rho}{\sqrt{3}}\left(\sigma_0+\frac{\sigma_8}{\sqrt{2}}\right),\label{vmf}
\end{equation}
which corresponds to the contribution from the mean-field approximation of
the nucleon field.
The one nucleon loop contribution $\Sigma_{ph}(\rho)$ is obtained as
\begin{equation}
 \Sigma_{ph}(\rho)=C_i\frac{g^2\rho}{m_N},\label{sph}
\end{equation}
where $i=\pi,\ \eta_0,\ \eta_8,\ \eta_0\eta_8$ and $C_\pi=\frac{1}{2},\
C_{\eta_0}=\frac{1}{3},\ C_{\eta_8}=\frac{1}{6},\
C_{\eta_0\eta_8}=\frac{1}{3\sqrt{2}}$.
These factor $C_i$ are obtained from the meson-baryon coupling constant
in the vacuum shown in Eq.(\ref{lagrangian}).
The contribution from the $\Sigma_{ph}(\rho)$ corresponds to the nucleon
particle-hole excitation.
The details of these calculations are shown in the Appendix \ref{appd}.
In the following, we assume that the nuclear matter does not contain
the strangeness component.

The value of the sigma condensate is
determined by minimizing the effective potential obtained from the linear
sigma model lagrangian.
As the result of the introduction of the medium effect, the effective
potential for $\sigma_0$ and $\sigma_8$ of the linear sigma model with the one loop
approximation is given as
\begin{eqnarray}
V_\sigma
 &=&\frac{{\mu}^2}{2}({\sigma}_0^2+{\sigma}_8^2)+\frac{\lambda}{12}({\sigma}_0^4+6{\sigma}_0^2{\sigma}_8^2-2\sqrt{2}{\sigma}_0{\sigma}_8^3+\frac{3}{2}{\sigma}_8^4)\nonumber\\
&&+\frac{{\lambda}^{\prime}}{4}({\sigma}_0^2+{\sigma}_8^2)^2
 -2Am_0{\sigma}_0-2Am_8{\sigma}_8\nonumber\\
&&-\frac{2}{3}B({\sigma}_0^3-\frac{3}{2}{\sigma}_0{\sigma}_8^2-\frac{{\sigma}_8^3}{\sqrt{2}})+\frac{g\rho}{\sqrt{3}}(\sigma_0+\frac{\sigma_8}{\sqrt{2}}),
\end{eqnarray}
where we have defined
\begin{eqnarray}
m_0&=&2m_q+m_s,\\
m_8&=&\sqrt{2}\left(m_q-m_s\right).
\end{eqnarray}
The term proportional to $g\rho$ comes from the medium effect from
the 1 loop diagram of the nucleon Eq.(\ref{vmf}).
If the nuclear density $\rho$ changes, the potential also changes.
Consequently, the vacuum, which is the minimum point of the potential,
changes.
The minimum conditions of the potential are given as
\begin{eqnarray}
\frac{\partial V_\sigma}{\partial
 {\sigma}_0}&=&{\mu}^2{\sigma}_0+\frac{\lambda}{6}(2{\sigma}_0^3+6{\sigma}_0{\sigma}_8^2-\sqrt{2}{\sigma}_8^3)+{\lambda}^{\prime}{\sigma}_0({\sigma}_0^2+{\sigma}_8^2)\nonumber\\
&&-2Am_0-2B({\sigma}_0^2-\frac{{\sigma}_8^2}{2})+\frac{g\rho}{\sqrt{3}}
 =0, \label{vc1}\\
 \frac{\partial V_\sigma}{\partial
 {\sigma}_8}&=&{\mu}^2{\sigma}_8+\lambda\sigma_8(\sigma_0^2-\frac{\sigma_0\sigma_8}{\sqrt{2}}+\frac{\sigma_8^2}{2})+{\lambda}^{\prime}{\sigma}_8({\sigma}_0^2+{\sigma}_8^2)\nonumber\\
&&-2Am_8+2B\sigma_8(\sigma_0+\frac{{\sigma}_8}{\sqrt{2}})+\frac{g\rho}{\sqrt{6}}
 =0. \label{vc2}
\end{eqnarray}
The solution for $\sigma_0=\left<\sigma_0\right>$ and $\sigma_8=\left<\sigma_8\right>$ of these equation is the
vacuum at non-zero $\rho$.
The in-medium meson masses are obtained from
\begin{equation}
 m^2(\rho)=m_0^2(\left<\sigma\right>^\ast)+\Sigma_{ph}(\rho).
\end{equation}
The first term $m_0^2(\left<\sigma\right>^\ast)$ is the same expression as in vacuum but evaluated with the
in-medium sigma condensate $\left<\sigma\right>^\ast$.
The in-vacuum meson masses are shown in Appendix \ref{appa}.
$m_0^2(\left<\sigma\right>^\ast)$ contains only the contribution from the diagram (A).
The contribution from the diagram (A) to the meson mass can be seen
explicitly by using the vacuum condition.
The in-medium masses of the pseudoscalar mesons $\pi, \eta_0, \eta_8$, which are
denoted as $m_\pi(\rho), m_{\eta_0}(\rho), m_{\eta_8}(\rho)$, and the mixing term of
$\eta_0$ and $\eta_8$, $m_{\eta_0\eta_8}^2(\rho)$, are following;
\begin{eqnarray}
m_{\pi}^2(\rho)&=&{\mu}^2+\frac{\lambda}{3}(\left<\sigma_0\right>^2+\sqrt{2}\left<\sigma_0\right>\left<\sigma_8\right>+\frac{\left<\sigma_8\right>^2}{2})\nonumber\\
&&+\lambda^\prime(\left<\sigma_0\right>+\left<\sigma_8\right>)\nonumber\\
&&-2B(\left<\sigma_0\right>-\sqrt{2}\left<\sigma_8\right>)+\frac{\sqrt{3}g\rho}{2\left(\left<\sigma_0\right>+\frac{\left<\sigma_8\right>}{\sqrt{2}}\right)}
 \nonumber\\
&=&\frac{6Am_q}{\left<\sigma_0\right>+\frac{\left< \sigma_8 \right>}{\sqrt{2}}}
\label{mpi}\\
m_{{\eta}_0}^2(\rho)&=&{\mu}^2+\frac{\lambda}{3}(\left<\sigma_0\right>^2+\left<\sigma_8\right>^2)+\lambda^\prime(\left<\sigma_0\right>^2+\left<\sigma_8\right>^2)\nonumber\\
&&+4B\left<\sigma_0\right>+\frac{g\rho}{\sqrt{3}\left(\left<\sigma_0\right>+\frac{\left<\sigma_8\right>}{\sqrt{2}}\right)}\nonumber\\
&=&6B\frac{\left(\left<\sigma_0\right>-\frac{\left<\sigma_8\right>}{\sqrt{2}}\right)^2}{\left<\sigma_0\right>-\sqrt{2}\left<\sigma_8\right>}\nonumber\\
&&+2A\left(\frac{2m_q}{\left<\sigma_0\right>+\frac{\left<\sigma_8\right>}{\sqrt{2}}}+\frac{m_s}{\left<\sigma_0\right>-\sqrt{2}\left<\sigma_8\right>}\right) \label{me0}\\
m_{\eta_8}^2(\rho)&=&\mu^2+\frac{\lambda}{3}(\left<\sigma_0\right>^2-\sqrt{2}\left<\sigma_0\right>\left<\sigma_8\right>+\frac{3\left<\sigma_8\right>^2}{2})\nonumber\\
&&+\lambda^\prime(\left<\sigma_0\right>^2+\left<\sigma_8\right>^2)\nonumber\\
&&-2B(\left<\sigma_0\right>+\sqrt{2}\left<\sigma_8\right>)+\frac{g\rho}{2\sqrt{3}\left(\left<\sigma_0\right>+\frac{\left<\sigma_8\right>}{\sqrt{2}}\right)}
 \nonumber\\
&=&6B\frac{\left<\sigma_8\right>^2}{\left<\sigma_0\right>-\sqrt{2}\left<\sigma_8\right>}\nonumber\\
&&+2A\left(\frac{m_q}{\left<\sigma_0\right>+\frac{\left<\sigma_8\right>}{\sqrt{2}}}+\frac{2m_s}{\left<\sigma_0\right>-\sqrt{2}\left<\sigma_8\right>}\right)\label{me8}\\
m_{\eta_0
 \eta_8}^{2}(\rho)&=&\frac{\sqrt{2}}{3}\lambda\left<\sigma_8\right>\left(\sqrt{2}\left<\sigma_0\right>-\frac{\left<\sigma_8\right>}{2}\right)-2B\left<\sigma_8\right>\nonumber\\
&&+\frac{g\rho}{\sqrt{6}\left(\left<\sigma_0\right>+\frac{\left<\sigma_8\right>}{\sqrt{2}}\right)}\nonumber\\
&=&-\frac{6B\left<\sigma_8\right>\left(\left<\sigma_0\right>-\frac{\left<\sigma_8\right>}{\sqrt{2}}\right)}{\left<\sigma_0\right>-\sqrt{2}\left<\sigma_8\right>}\nonumber\\
&&+2\sqrt{2}A\left(\frac{m_q}{\left<\sigma_0\right>+\frac{\left<\sigma_8\right>}{\sqrt{2}}}-\frac{m_s}{\left<\sigma_0\right>-\sqrt{2}\left<\sigma_8\right>}\right)\nonumber\\
 \label{me08}
\end{eqnarray}
Here we have used the vacuum condition Eq.(\ref{vc1}) and Eq.(\ref{vc2}) to obtain the second expressions.
It is interesting that in the second expressions for the in-medium meson masses the explicit density dependence disappears.
This is a consequence of chiral symmetry in meson-nucleon interaction, in which the sigma exchange and Born contributions are cancelled away.
\begin{figure}[t]
\includegraphics[width=9cm,clip]{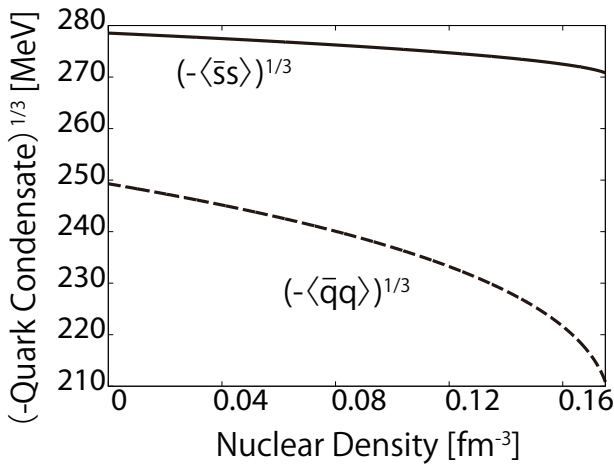}
\caption{\label{qcdd1}The chiral condensates in the nuclear medium. The dashed and solid lines denote the $(-\left<\bar{q}q\right>)^{1/3}$
 and $(-\left<\bar{s}s\right>)^{1/3}$, respectively.}
\end{figure}
The physical masses of $\eta$ and $\eta^\prime$ are obtained by
\begin{eqnarray}
m_{\eta}^2&=&\frac{1}{2}\left(
			 m_{{\eta}_0}^2+m_{{\eta}_8}^2-\sqrt{(m_{{\eta}_0}^2-m_{{\eta}_8}^2)^2+4{m_{08}^P}^4})
			\right) \nonumber\\ \\
m_{{\eta}^{\prime}}^2&=&\frac{1}{2}\left(
				    m_{{\eta}_0}^2+m_{{\eta}_8}^2+\sqrt{(m_{{\eta}_0}^2-m_{{\eta}_8}^2)^2+4{m_{08}^P}^4})
				   \right)\nonumber\\ 
\end{eqnarray}
so as to resolve the off diagonal mass term $m_{\eta_0\eta_8}^2$.

From these explicit forms of the $\eta_0$ and $\eta_8$ meson mass, the
mass difference of these mesons in the SU(3) flavor symmetric limit
($m_q=m_s$, $\left<\sigma_8\right>=0$) is written as
 \begin{equation}
  m_{\eta_0}^2-m_{\eta_8}^2=6B\left<\sigma_0\right>. \label{massdiff}
 \end{equation}
This is the consistent expression to the discussion in Sec.\ref{massred1}, where
we have shown both effects of the U$_A$(1) anomaly and the chiral
symmetry breaking are necessary for the mass difference of $\eta_0$ and
$\eta_8$.
In addition, since $\eta_8$ is the Nambu-Goldstone boson associated with the spontaneous breaking of chiral SU(3) symmetry, the mass of the $\eta_8$ meson comes from the explicit breaking of chiral symmetry.
Assuming that the $\eta_0$ and $\eta_8$ masses are orders of 1000MeV and 500MeV, respectively, one finds from Eq.(\ref{massdiff}) that almost half of the $\eta_0$ mass is generated by the spontaneous chiral symmetry breaking through the U$_A$(1) anomaly.


In the following, we show the in-medium meson masses calculated with the
medium effect including the SU(3)$_V$ breaking owing to the quark mass
difference.
The parameters are determined by the method shown in Appendix \ref{appb}.
As the input parameters, we used $f_\pi, f_K, m_\pi, m_K, m_\sigma$, the
sum of $m_\eta^2$ and $m_{\eta^\prime}^2$, and the degenerate $u$, $d$ quark
mass $m_q$. 
All the used and determined parameters are shown in Appendix \ref{appb}.
We determine the meson-baryon coupling parameter $g$ by
the reduction of the chiral condensate.

First, we show the density dependence of the chiral condensate in
Fig.\ref{qcdd1}.
Since the parameter $g$ is determined to reproduce the 35\% reduction of the quark condensate at normal
nuclear density, the quark condensate at the saturation density is the input value here.
As mentioned above, we assume that the nuclear medium contains no explicit
strange component.
So, the strange condensate is insensitive to the nuclear density.
Nevertheless, the strange condensate does change slightly through the
SU(3) flavor breaking of the nuclear matter.

Next, we show the result of the in-medium meson masses including the SU(3)
breaking by the current quark mass in Fig.\ref{mmdd1}.
\begin{figure}
\includegraphics[width=8cm,clip]{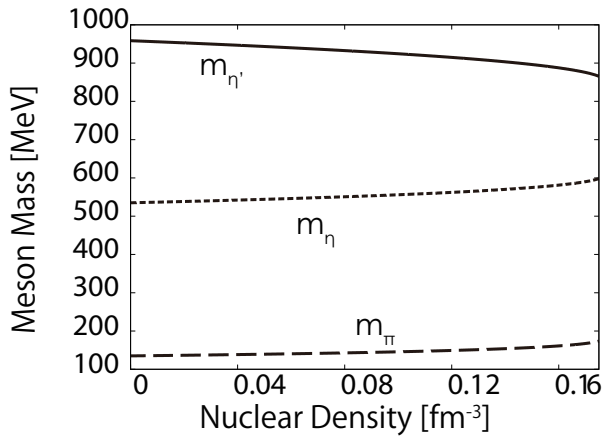}
 \caption{\label{mmdd1}The mass shift of $\eta^\prime$ meson in the
  nuclear medium. The solid, dotted and dashed lines represent the $\eta^\prime$, $\eta$ and $\pi$ meson masses in the nuclear medium, respectively.}
\end{figure}
From this calculation, we find that the $\eta^\prime$ mass reduces about
80MeV at the normal nuclear density.
In contrast, the masses of the other pseudoscalar octet mesons are enhanced.
Especially for the $\eta$ case, the enhancement is about 50MeV.
This is because under the partial restoration of chiral symmetry the magnitude
of the spontaneous breaking is suppressed and consequently the
Nambu-Goldstone boson nature of the octet pseudoscalar mesons declines.

Finally, we show the density dependence of the mixing angle of
$\eta_0$-$\eta_8$ in Fig.\ref{ma}.
we defined the mixing angle $\theta$ with
\begin{equation}
\tan 2\theta=\frac{2m_{\eta_0\eta_8}^2}{m_{\eta_0}^2-m_{\eta_8}^2}
 \label{mixingangle}.
\end{equation}
The density dependence of the mixing angle $\theta$ is shown in Fig.\ref{ma}.
\begin{figure}[t]
\includegraphics[width=7cm,clip]{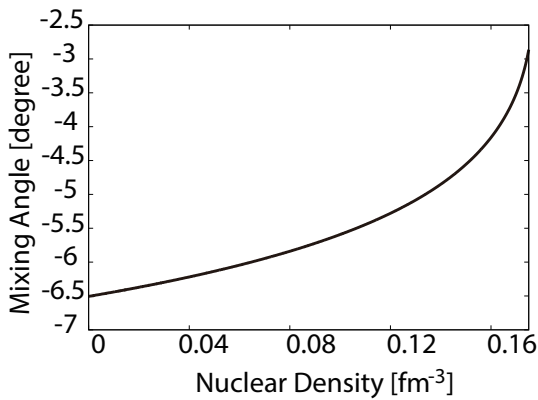}
 \caption{The $\eta_0$-$\eta_8$ mixing angle in the nuclear medium}
\label{ma}
\end{figure}
As we can see in Fig.\ref{ma}, the absolute value of the mixing angle
becomes smaller when the nuclear density become larger.
One can understand this behavior as follows;
When chiral symmetry is being restored partially with the reduction of the magnitude of the sigma condensates, the first terms of Eq.(\ref{me0}), (\ref{me8}) and (\ref{me08}) are getting suppressed.
In the limit where the first terms vanish, the mixing angle is obtained by $\tan 2\theta=2\sqrt{2}$ and has a positive large value.
Therefore, the mixing angle is approaching to the positive value with the partial restoration.

\section{\label{sec:epnint}The low energy \texorpdfstring{$\eta^\prime N$}{eta'N} interaction in vacuum}
Let us discuss the $\eta^\prime N$ two-body interaction in vacuum.
In the following, we estimate the $\eta^\prime N$ interaction strength with the
linear sigma model developed in the previous section.
We evaluate the invariant amplitude of the meson and nucleon $V_{ab}$ in
the tree level by the scalar meson exchange and Born terms shown in
Fig.\ref{etaNint}:
\begin{eqnarray}
 &&-iV_{ab}=g_{\sigma_0
  NN}C^{(0)}_{ab}\frac{i}{(k-k^\prime)^2-m_{\sigma_0}^2}\nonumber\\
&&+g_{\sigma_8 NN}C^{(8)}_{ab}\frac{i}{(k-k^\prime)^2-m_{\sigma_8}^2}\nonumber\\
&&\ +C_a\gamma_5 \frac{i}{\Slash{p}+\Slash{k}-m_N}C_b \gamma_5 +C_b\gamma_5 \frac{i}{\Slash{p}-\Slash{k^\prime}-m_N}C_a\gamma_5,\nonumber\\ \label{invamp}
\end{eqnarray}
where $k$ and $k^\prime$ are in-coming and out-going meson momenta,
respectively, and $p$ is the in-coming nucleon momentum.
The labels $a, b$ correspond to the in-coming and out-going mesons,
$C_{ab}^{(0)}$, $C_{ab}^{(8)}$ are the coupling constant of the pseudoscalar
mesons and $\sigma_0$ and $\sigma_8$ meson and $g_{\sigma_0 NN}, g_{\sigma_8 NN}$ are $\sigma_0$, $\sigma_8$ and nucleon coupling, respectively, and $C_a$ is the coupling constant between the pseudoscalar meson and the nucleon.
The first term is the contribution from the scalar meson exchange shown
in diagram (a) of Fig.\ref{etaNint}, while the second and third terms are
the Born terms shown as diagram (b) and (c) of Fig.\ref{etaNint}.

With the meson momentum expansion, the Lorentz scalar part of the sum of
the amplitude for the NG boson and nucleon scattering is cancelled out,
while the vector part remains the contribution.
This interaction is known as the Weinberg-Tomozawa (WT) low energy
theorem stemming from the spontaneous chiral symmetry breaking.
In the flavor SU(3) limit and $a,b\neq \eta_0$,
the vacuum condition in the chiral limit is given as
\begin{equation}
 \mu^2+\frac{\lambda}{3}\left<\sigma_0\right>^2+\lambda^\prime\left<\sigma_0\right>^2-2B\left<\sigma_0\right>=0,\label{vcond}
\end{equation}
the scalar and pseudoscalar meson coupling are
\begin{eqnarray}
 C_{ab}^{(0)}&=&-i\delta^{ab}\left(\frac{2}{3}\lambda\left<\sigma_0\right>+2\lambda^\prime\left<\sigma_0\right>-2B\right),\label{cab0}\\
 C_{ab}^{(8)}&=&-i\delta^{ab}\left(\frac{\sqrt{2}}{3}\lambda\left<\sigma_0\right>+2\sqrt{2}B\right).
\end{eqnarray}
$\sigma_0$ and $\sigma_8$ and nucleon coupling are
\begin{eqnarray}
g_{\sigma_0 NN}&=&-i\frac{g}{\sqrt{3}}\\
g_{\sigma_8 NN}&=&-i\frac{g}{\sqrt{6}}
\end{eqnarray}
The meson-baryon coupling $C_a$ is given as
\begin{equation}
 C_{a}=\frac{g}{\sqrt{2}}\tau_a\ \ (a=1,2,3,8)
\end{equation}
from Eq.(\ref{lagrangian}).
Here, we define $\tau_8\equiv \frac{1}{\sqrt{3}}\cdot {\bf 1}$.
The masses are
\begin{eqnarray}
m_{\sigma_0}^2&=&\mu^2+\lambda\left<\sigma_0\right>^2+3\lambda^\prime\left<\sigma_0\right>^2-4B\left<\sigma_0\right>\nonumber\\
&=&\frac{2}{3}\lambda\left<\sigma_0\right>^2+2\lambda^\prime\left<\sigma_0\right>^2-2B\left<\sigma_0\right>,\label{chilims0}\\
m_{\sigma_8}^2&=&\mu^2+\lambda\left<\sigma_0\right>^2+\lambda^\prime\left<\sigma_0\right>^2+2B\left<\sigma_0\right>\nonumber\\
&=&\frac{2}{3}\lambda\left<\sigma_0\right>^2+4B\left<\sigma_0\right>,\label{chilims8}\\
m_N&=&\frac{g}{\sqrt{3}}\left<\sigma_0\right>\label{chilimmn},
\end{eqnarray}
where we used the vacuum condition Eq.(\ref{vcond}).
Substituting Eq.(\ref{cab0}-\ref{chilimmn}) for Eq.(\ref{invamp}) and
expanding the amplitude in terms of the in-coming and out-going meson momenta $k, k^\prime$, we can obtain the $s$-wave
amplitude of the NG boson ($a,\ b\neq 0$) and baryon scattering as
\begin{equation}
 V_{ab}= -\frac{g^2\omega}{8m_N^2}\left[ \tau_a, \tau_b \right]
+\mathcal{O}(k^2),\label{wt}
\end{equation}
where $\omega$ is the meson energy.
Here, we have used the Dirac equation $(\Slash{p}-m_N)u(p)=0$ and we
take only the $s$-wave contribution for low energy scattering.
In Eq.(\ref{wt}), we omitted the unit matrix of the spinor space.
\begin{figure}[t]
\includegraphics[width=9cm,clip]{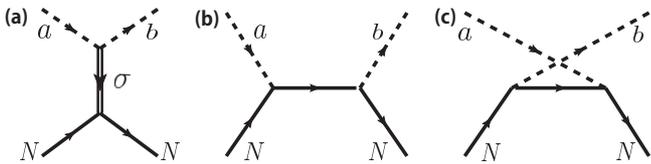}
\caption{\label{etaNint}The diagrams which contribute the $\eta^\prime N$ interaction. The dashed, single, double lines mean the pseudoscalar
 meson, nucleon, scalar meson propagation, respectively.}
\end{figure}

In the case of the $\eta^\prime N$ interaction, the interaction strength
in the chiral limit is derived as follows.
From Eq.(\ref{lagrangian}), the $\eta_0\sigma_0$ coupling and the $\eta_0\sigma_8$ coupling in the SU(3) symmetric limit can be obtained from the lagrangian Eq.(\ref{lagrangian}) as
\begin{eqnarray}
C_{\eta_0 \eta_0}^{(0)}&=&-i(\frac{2}{3}\lambda\left<\sigma_0\right>+2\lambda^\prime\left<\sigma_0\right>+4B),\\
C_{\eta_0\eta_0}^{(8)}&=&0,
\end{eqnarray}
and $\eta_0$ and nucleon coupling $C_{\eta_0 N}$ can read
\begin{equation}
 C_{\eta_0 N}=\frac{g}{\sqrt{3}}\cdot {\bf 1}.
\end{equation}
Substituting $C_{\eta_0 \eta_0}^{(0,8)}$ and $C_{\eta_0 N}$ for Eq.(\ref{invamp}), the $\eta_0 N$ interaction in the linear sigma model
in the chiral limit and low energy compared to the meson and nucleon
mass is calculated as
\begin{eqnarray}
-iV_{\eta_0N}&=&
-\frac{ig}{\sqrt{3}}i(\frac{2}{3}\lambda\left<\sigma_0\right>+2\lambda^\prime\left<\sigma_0\right>+4B)\frac{i}{m_{\sigma_0}^2}\nonumber\\
&&+\left(\frac{g}{\sqrt{3}}\right)^2\gamma_5\left(\frac{i}{\Slash{p}+\Slash{k}-m_N}+\frac{i}{\Slash{p}-\Slash{k}-m_N}\right)\gamma_5\nonumber\\
&=&\frac{ig}{\sqrt{3}\left<\sigma_0\right>}\frac{m_{\sigma_0}^2+6B\left<\sigma_0\right>}{m_{\sigma_0}^2}\nonumber\\
&&-\frac{i}{2}\left(\frac{g}{\sqrt{3}}\right)^2\left(\frac{\Slash{k}}{p\cdot k}+\frac{\Slash{k}^\prime}{p\cdot k^\prime}\right)+\mathcal{O}(k^2)\nonumber\\
 &=&\frac{ig^2}{3m_N}\left(1+\frac{6B\left<\sigma_0\right>}{m_{\sigma_0}^2}\right)-i\frac{g^2}{3m_N}+\mathcal{O}(k^2)\nonumber\\
 &=&\frac{ig}{\sqrt{3}}\frac{6B}{m_{\sigma_0}^2}+\mathcal{O}(k^2)\label{etanint}
\end{eqnarray}
From the first line to the second line, we have kept the leading contribution
in the meson momentum expansion and replaced $\Slash{p}$ with $m_N$ as well as the case of the NG boson and nucleon
scattering, and from the third line to the fourth line we take only
$s$-wave amplitude for the low energy scattering.
As a result, the leading contribution to the $\eta_0 N$ interaction is induced by the $B$
term, which comes from the U$_A$(1) anomaly.
In contrast, the Weinberg-Tomozawa interaction is cancelled due to the U$_A$(1) symmetry.
This is because only the terms including $B$ (and the quark mass) break the U$_A$(1) chiral symmetry and the other terms keep the symmetry.
Thanks to the chiral symmetry in these terms, we have the cancellation between the $\sigma$ exchange and Born terms.

Substituting the values of the parameters into Eq.(\ref{etanint}),
we find the $\eta^\prime N$ interaction to be attractive with strength
-0.0534MeV$^{-1}$.
This attraction is strong comparable to the $\bar{K}N$ system with $I=0$, in
which it is conceivable that there exists a $\bar{K}N$ quasi-bound state
regarded as $\Lambda(1405)$.
In the following, we use this value as the $\eta^\prime N$ coupling constant.

\section{\label{epnbs1}The \texorpdfstring{$\eta^\prime N$}{eta'N} bound state}
In the previous section, we have obtained the tree-level amplitude for
the $\eta^\prime N$ scattering in the linear sigma model.
Making use of this amplitude as an interaction kernel, we solve a
scattering equation for the $\eta^\prime N$ two-body system.
Because the $\eta^\prime N$ interaction is attractive with a comparable
strength to the $\bar{K}N$ system with $I=0$, we expect that the
$\eta^\prime N$ system forms a bound state similar to $\Lambda(1405)$,
which is a bound state in the $\bar{K}N$ channel.
In this section, we evaluate the scattering length of the
$\eta^\prime N$ system and binding energy if an $\eta^\prime N$ bound
state is formed.

To solve the $\eta^\prime N$ scattering system, we make use of the same
machinery for the $\Lambda(1405)$ in the $\bar{K}N$ channel with $I=0$
\cite{Jido1998av}, in which the $\bar{K}N$ scattering amplitude obtained
with the chiral perturbation theory at the tree level is used as the
interaction kernel of the scattering equation and the loop function is
regularized so that the scattering amplitude can be described in terms
of hadronic objects.
As a result one finds a quasi-bound state in
the $\bar{K}N$ channel.
The $T$-matrix is calculated by the single-channel Lippmann-Schwinger equation.
Here we take the $\eta^\prime N$ interaction evaluated in
Eq.(\ref{etanint}) as the interaction kernel.
We denote the interaction kernel as $V_{kk'}$, where the indices $k$ and $k'$ are in-coming and out-going meson momenta respectively.
Now, we are in the case that the interaction kernel $V_{kk'}$ is independent
of the external momentum, the $T$-matrix can be obtained in an algebraic way;
\begin{eqnarray}
 T_{kk'}&=&V_{kk'}+\int dlV_{kl}G_{l}T_{lk'} \nonumber\\
 &=&V_{kk'}+\int dlV_{kl}G_l V_{lk'}\nonumber\\
&&+\int dl\int dl' V_{kl}G_lV_{ll'}G_{l'}V_{l'k'}+\dots \nonumber\\
 &=&V\sum_{n=0}^\infty\left(V\int dlG_l\right)^n= \frac{V}{1-V\int dlG_l},\label{lseq}
\end{eqnarray}
where $G_l$ is the two-body Green function of $\eta^\prime$ and nucleon.
From the second line to the third line, we used the fact that the
interaction kernel $V_{kk'}$ is independent of the external momentum, $V_{kk'}=V$.
Because we take the momentum-independent contact interaction
Eq.(\ref{etanint}), the integral of $G_l$ diverges.
With dimensional regularization, the integral of $G_l$ is calculated with
\begin{eqnarray}
G(W) &\equiv&\int dl G_l\nonumber\\
&=& i \int \frac{d^4l}{(2\pi)^4} \frac{2m_N}{l^2-m_N^2+i\epsilon} \frac{1}{(P-l)^2-m_{\eta^\prime}^2+i\epsilon}  \nonumber\\
&=& \frac{2m_N}{16{\pi}^2} \{ a(\mu)+\ln \frac{m_N^2}{{\mu}^2}+\frac{m_{\eta^\prime}^2-m_N^2+W^2}{2W^2}\ln \frac{m_{\eta^\prime}^2}{m_N^2} \nonumber\\
&&+\frac{\bar{q}}{W}[\ln (W^2-(m_N^2-m_{\eta^\prime}^2)+2\bar{q}W)\nonumber\\
&&+\ln (W^2+(m_N^2-m_{\eta^\prime}^2)+2\bar{q}W)\nonumber\\
&&-\ln (-W^2-(m_N^2-m_{\eta^\prime}^2)+2\bar{q}W)\nonumber\\
&&-\ln (-W^2+(m_N^2-m_{\eta^\prime}^2)+2\bar{q}W)]\},\label{loop}
\end{eqnarray}
where $\mu$ is the scale of dimensional regularization and the
center-of-mass momentum is given by
\begin{equation}
 \bar{q} = \frac{\sqrt{(W^2-(m_N+m_{\eta^\prime})^2)(W^2-(m_N-m_{\eta^\prime})^2)}}{2W}
\end{equation}
From the second line to the third line of Eq.(\ref{loop}), we have
supposed that the divergent part of could be absorbed to interaction
vertices in the renormalization procedure and the remaining finite
constant is denoted as $a(\mu)$.
The subtraction constant $a(\mu)$ has to be determined in some way.
Here we take the natural renormalization scheme proposed in
\cite{Jido1998av} in which the CDD pole contribution are excluded from
the scattering amplitude in a consistent way with chiral counting.
This means that the scattering amplitude is described by dynamics of
$\eta^\prime$ and $N$.
In our calculation, we use $a(\mu)=-1.838$ and the renormalization point
$\mu=m_N$.

Using the $T$ matrix calculated with the above method, we evaluate the
binding energy and scattering length of the $\eta^\prime N$ system.
The mass $m_B$ of the bound state is obtained as the pole position of
the $T$-matrix.
The binding energy $E_B$ is calculated by $E_B=m_N+m_{\eta^\prime
N}-m_B$.



With Eqs.(\ref{lseq}) and (\ref{loop}), the $\eta^\prime N$ binding
energy $E_B$ is obtained as 6.2MeV.
The scattering length and effective range are obtained as -2.7fm and
0.25fm with the definition in Appendix \ref{appe}.
We show the scattering amplitude with $m_{\sigma_0}=700$MeV in
Fig.\ref{sa1}.
\begin{figure}[t]
 \includegraphics[width=8cm,clip]{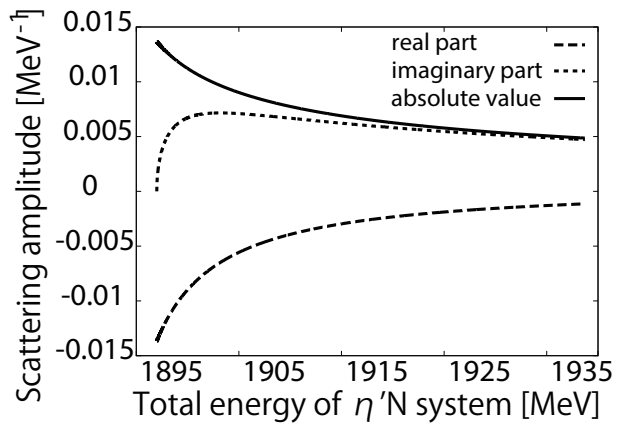}
\caption{\label{sa1}The value of the scattering amplitude above the threshold. The dashed, dotted and solid lines represent the real part,
 the imaginary part and the absolute value of the scattering amplitude of
 the $\eta^\prime N$ system with $m_{\sigma_0}=700$MeV, respectively.
}
\end{figure}

In this calculation, we used the mass of the sigma meson $m_{\sigma_0}$
as an input to fix the parameter of the lagrangian of the linear sigma model.
In the previous calculations, we used the $m_{\sigma_0}=700$MeV.
The sigma meson mass dependence of the binding
energy, scattering length and effective range is given in TABLE \ref{betable}.
The parameters of the lagrangian are determined for each $m_{\sigma_0}$
with the procedure shown in Sec.\ref{med1}.
Within the wide range of the $m_{\sigma_0}$, we found the existence of
the $\eta^\prime N$ bound state and the binding energy have a somewhat
$m_{\sigma_0}$ dependence.
From the TABLE \ref{betable}, we find that the larger binding energy
accompanies the smaller scattering length.
This behavior can be understood because the scattering length can be roughly
evaluated with  $1/\sqrt{2mE_B}$, where $m$ is
the reduced mass of $\eta^\prime$ and nucleon.
We find the scattering length is about 1fm if a $\eta^\prime N$ bound
state exists with the binding energy a few MeV. 
\begin{table}[t]
\caption{The $m_{\sigma_0}$ dependence of the $\eta^\prime N$ bound state}
\label{betable}
\begin{center}
\setlength{\tabcolsep}{3pt}
\footnotesize
\begin{tabular}{c|c|c|c}\hline
 $m_{\sigma_0}$&binding energy&scattering length&effective
 range \\
 $[$MeV]&[MeV]&[fm]&[fm]\\ \hline
500 & 3.5 &-3.5&0.25\\
600 & 6.2 &-2.7&0.25\\
700 & 6.2 &-2.7&0.25\\
800 & 4.6 &-3.1&0.25\\
900 & 2.4 &-4.2&0.26\\
1000 & 0.6 &-8.1&0.33\\ \hline
\end{tabular}
\end{center}
\end{table}

The result in the low energy limit depends on the choice of the
subtraction constant $a(\mu)$ and we determined $a(\mu)$ to exclude
other dynamics than $\eta^\prime$ and $N$ here.
The other degree of freedom, for example the $\omega$ meson exchange
interaction or the microscopic quark dynamics, may spoil such a description
\cite{Hyodo2008}.

\section{\label{conc}Conclusion and Remark}
In this paper, we have constructed a chiral effective lagrangian for the
mesons based on the linear realization of the SU(3) chiral symmetry in the symmetric nuclear matter and estimated the mass reduction of $\eta^\prime$ in the medium.
The lagrangian contains the explicit
 breaking of the chiral symmetry and flavor symmetries and the determinant type U$_A$(1) breaking term which introduces the effect of the U$_A$(1) anomaly.
 We find that a substantial part of the $\eta^\prime$ mass is generated by the spontaneous breaking of chiral symmetry through the U$_A$(1) anomaly.
The nuclear matter is taken into account as a mean field by calculating one nucleon loop in the Fermi gas.
The parameters of the lagrangian have been fixed by the observed quantities, such as the meson decay constants and the meson masses.
In the determination of the coupling strength of nucleon and the sigma meson, we make use of partial restoration of chiral symmetry, that is, the experimental suggestion
of the 30\% reduction of the quark condensate as the basic assumption.
In our calculation, we have obtained the 80MeV reduction of the
$\eta^\prime$ meson mass at the normal nuclear density.

Based on the effective lagrangian used for the calculation of the in-medium properties of the mesons, we have also estimated the 2-body $\eta^\prime N$ interaction in vacuum.
Using the interaction of $\eta^\prime N$ as the kernel of the scattering equation, we have evaluated the $T$-matrix of the $\eta^\prime N$ system.
As the result, we have obtained a $\eta^\prime N$ bound state, which is
an analogous state of $\Lambda(1405)$ in the $\bar{K}N$ system.
The binding energy of the system is found to be several MeV, which is comparable to the typical value of
the hadronic bound state, for example, $\Lambda(1405)$ or deuteron.
We have also evaluated the scattering length and the effective range of the $\eta^\prime N$ system,
having obtained a few fm with the repulsive sign for the scattering length, which is a consequence of the existence of the bound state, and a quarter fm of the effective range.

In the linear sigma model, the $\eta^\prime N$ interaction is originated
from the sigma meson exchange with the $\eta_0\eta_0\sigma$ coupling coming from the U$_A$(1) breaking determinant term.
The Weinberg-Tomozawa type vector interaction is cancelled away by the scalar-meson-exchange and Born terms thanks to chiral symmetry.
In contrast, the interactions of the octet pseudoscalar meson and nucleon are expressed by the Weinberg-Tomozawa interaction at low energies as a consequence of the spontaneous breaking of chiral symmetry, and there is no sigma exchange term, which is cancelled away with the Born term and turns into the Weinberg-Tomozawa interaction.
This implies that the difference comes from the fact that the $\eta^\prime$ meson is not a Nambu-Goldstone boson due to the U$_A$(1) anomaly.

Actually, the $\sigma\eta_0\eta_0$ coupling from the explicit U$_A$(1) breaking induces the mass of the $\eta^\prime$ meson when chiral symmetry is broken spontaneously with finite $\sigma$ condensate.
In this way, the $\sigma\eta_0\eta_0$ coupling plays an important role for the mass generation of the $\eta^\prime$ meson.
This is the case also for nucleon.
The nucleon mass is generated by the sigma condensate through the $\sigma NN$ coupling.
In addition, the strong $\sigma NN$ coupling induces a strong attraction in the scalar-isoscalar channel for the $NN$ interaction with the $\sigma$ meson exchange, 
Thus, we conclude that the $\eta^\prime N$ interaction in the scalar channel entirely analogous to the $NN$ interaction.
Since the $\sigma\eta_0\eta_0$ and $\sigma NN$ coupling are necessary for the mass generation of the $\eta^\prime$meson and nucleon in the linear sigma model, the $\eta^\prime N$ interaction coming from the $\sigma$ exchange is inevitable.
(In the same manner, one could have a strong attraction also in the $\eta^\prime \eta^\prime$ system.)
This attraction may open the possibility to have bound states in $\eta^\prime N$ and $\eta^\prime$-nucleus systems.
Nevertheless, there could be such repulsive interactions in other channels as to spoil the bound states.
It should be noted that chiral symmetry says that there is no Weinberg-Tomozawa interaction in the $\eta_0 N$ channel.

\begin{acknowledgements}
S. S. appreciates the support by the Grant-in-Aid for JSPS Fellows (No. 25-1879).
This work was partially supported by the Grants-in-Aid for Scientific Research 
(No. 25400254 and No. 24540274).
\end{acknowledgements}

\appendix
\section{\label{appa}The linear sigma model in vacuum}
In this section, we show the application of the linear sigma model in vacuum.
From the meson lagrangian Eq.(\ref{mesonl}), we obtain the effective
potential for $\sigma_0$ and $\sigma_8$, $V_\sigma(\sigma_0,\sigma_8)$, using tree approximation as follows;
\begin{eqnarray}
 V_\sigma(\sigma_0,\sigma_8)&=&\frac{\mu^2}{2}(\sigma_0^2+\sigma_8^2)\nonumber\\
&&+\frac{\lambda}{12}(\sigma_0^4+6\sigma_0^2\sigma_8^2-2\sqrt{2}\sigma_0\sigma_8^3+\frac{3}{2}\sigma_8^4)\nonumber\\
&&+\frac{\lambda^\prime}{4}(\sigma_0^2+\sigma_8^2)^2-2A(m_0\sigma_0+m_8\sigma_8)\nonumber\\
&&-\frac{2}{3}B(\sigma_0^3-\frac{3}{2}\sigma_0\sigma_8^2-\frac{\sigma_8^3}{\sqrt{2}})\label{vsigma}
\end{eqnarray}
In Eq.(\ref{vsigma}), we have omitted $\sigma_3$ because we assume the isospin symmetry and trivially $\left<\sigma_3\right>=0$.
The vacuum expectation values, $\left<\sigma_0\right>, \left<\sigma_8\right>$, are obtained as the
minimum point of the potential.
The minimum point is obtained by solving the following vacuum condition;
\begin{eqnarray}
 \frac{\partial V_\sigma}{\partial
 {\sigma}_0}&=&{\mu}^2{\sigma}_0+\frac{\lambda}{6}(2{\sigma}_0^3+6{\sigma}_0{\sigma}_8^2-\sqrt{2}{\sigma}_8^3)+{\lambda}^{\prime}{\sigma}_0({\sigma}_0^2+{\sigma}_8^2)\nonumber\\
&&-2Am_0-2B({\sigma}_0^2-\frac{{\sigma}_8^2}{2})
 =0 \label{vc3}\\
 \frac{\partial V_\sigma}{\partial
  {\sigma}_8}&=&{\mu}^2{\sigma}_8+\frac{\lambda}{2}(2{\sigma}_0^2{\sigma}_8-\sqrt{2}{\sigma}_0{\sigma}_8^2+{\sigma}_8^3)+{\lambda}^{\prime}{\sigma}_8({\sigma}_0^2+{\sigma}_8^2)\nonumber\\
&&-2Am_8+2B({\sigma}_0{\sigma}_8+\frac{{\sigma}_8^2}{\sqrt{2}}) =0, \label{vc4}
\end{eqnarray}
where, we have defined
\begin{eqnarray}
m_0&=&2m_q+m_s \\
m_8&=&\sqrt{2}(m_q-m_s).
\end{eqnarray}

The meson masses are obtained as the second order derivative of the full effective potential $V$ at the vacuum point $\frac{\partial V_\sigma}{\partial \sigma}=0$;
\begin{eqnarray}
 \left. m_{ab}^2=\frac{\partial^2V}{\partial\pi^a\partial\pi^b}\right|_{\pi^{a,b}=0}.
\end{eqnarray}
Here, $\pi^a$ is the meson field and $m_a\equiv m_{aa}$ stands for the
mass of the meson $\pi^a$ and $m_{ab}$ ($a\neq b$) means the mixing term
between $\pi^a$ and $\pi^b$.
Using the vacuum expectation values,
$\left<\sigma_0\right>, \left<\sigma_8\right>$,
we obtain the meson masses as follows;
\begin{eqnarray}
m_{{\sigma}_0}^2&=&{\mu}^2+\lambda(\left<{\sigma}_0\right>^2+\left<{\sigma}_8\right>^2)\nonumber\\
&&+{\lambda}^{\prime}(3\left<{\sigma}_0\right>^2+\left<{\sigma}_8\right>^2)-4B\left<{\sigma}_0\right> \label{ms0}\\
&=&\frac{2}{3}\lambda(\left<\sigma_0\right>^2+\frac{\left<\sigma_8\right>^3}{2\sqrt{2}\left<\sigma_0\right>})+2\lambda^\prime\left<\sigma_0\right>^2\nonumber\\
&&-2B(\left<\sigma_0\right>+\frac{\left<\sigma_8\right>^2}{2\left<\sigma_0\right>})+\frac{2Am_0}{\left<\sigma_0\right>}\label{ms0v1}\\
m_{{\sigma}_8}^2&=&{\mu}^2+\lambda(\left<{\sigma}_0\right>^2-\sqrt{2}\left<{\sigma}_0\right>\left<{\sigma}_8\right>+3\left<{\sigma}_8\right>^2/2) \nonumber\\
&&+{\lambda}^{\prime}(\left<{\sigma}_0\right>^2+3\left<{\sigma}_8\right>^2 )+2B({\sigma}_0+\sqrt{2}{\sigma}_8) \\
&=&\lambda(\frac{2}{3}\left<\sigma\right>^2-\sqrt{2}\left<\sigma_0\right>\left<\sigma_8\right>+\frac{\left<\sigma_8\right>^3}{3\sqrt{2}\left<\sigma_0\right>}+\frac{\left<\sigma_8\right>^2}{2})\nonumber\\
&&+2\lambda^\prime\left<\sigma_8\right>^2+\frac{2Am_0}{\left<\sigma_0\right>}\nonumber\\
&&+B(4\left<\sigma_0\right>+2\sqrt{2}\left<\sigma_8\right>-\frac{\left<\sigma_8\right>^2}{\left<\sigma_0\right>})\\
m_{\sigma_0\sigma_8}^2&=&\frac{\lambda}{2}(4{\sigma}_0{\sigma}_8-\sqrt{2}{\sigma}_8^2)+2{\lambda}^{\prime}{\sigma}_0{\sigma}_8+2B{\sigma}_8 \\
m_\pi^2 &=& \mu^2+\frac{\lambda}{3} ( \left<\sigma_0\right>^2+\sqrt{2}\left< \sigma_0 \right> \left< \sigma_8 \right>+\left< \sigma_8 \right>^2/2 ) \nonumber \\
&&+\lambda^\prime (\left<\sigma_0\right>^2+\left<\sigma_8\right>^2)-2B(\sigma_0-\sqrt{2}\sigma_8) \label{mpiv4}\\
&=&\frac{2\sqrt{6}Am_q}{f_\pi}\label{mpiv2}\\
m_K^2 &=& \mu^2+\frac{\lambda}{3}(\left<{\sigma}_0\right>^2-\left<{\sigma}_0\right>\left<{\sigma}_8\right>/\sqrt{2}+7\left<{\sigma}_8\right>^2/2)\nonumber\\
&&+{\lambda}^{\prime}(\left<{\sigma}_0\right>^2+\left<{\sigma}_8\right>^2)-2B(\left<{\sigma}_0\right>+\left<{\sigma}_8\right>/\sqrt{2}) \label{mkv4}\\
\label{mkv1}\\
&=&\frac{\sqrt{6}A\left(m_q+m_s\right)}{f_K} \label{mkv2}\\
m_{{\eta}_0}^2&=&{\mu}^2+\frac{\lambda}{3}(\left<{\sigma}_0\right>^2+\left<{\sigma}_8\right>^2)\nonumber\\
&&+{\lambda}^{\prime}(\left<{\sigma}_0\right>^2+\left<{\sigma}_8\right>^2)+4B\left<{\sigma}_0\right>\label{me0v1}\\
&=&\sqrt{\frac{2}{3}}B\frac{(4f_K-f_\pi)^2}{2f_K-f_\pi}\nonumber\\
&&+\frac{2\sqrt{2}}{\sqrt{3}}A\left( \frac{2m_q}{f_\pi}+\frac{m_s}{2f_K-f_\pi} \right)\label{me0v2}\\
m_{{\eta}_8}^2&=&{\mu}^2+\frac{\lambda}{3}(\left<{\sigma}_0\right>^2-\sqrt{2}\left<{\sigma}_0\right>\left<{\sigma}_8\right>+3\left<{\sigma}_8\right>^2/2)\nonumber\\
&&+{\lambda}^{\prime}(\left<{\sigma}_0\right>^2+\left<{\sigma}_8\right>^2)-2B(\left<{\sigma}_0\right>+\sqrt{2}\left<{\sigma}_8\right>)\nonumber\\
\label{me8v1}\\
&=&\frac{8\sqrt{2}}{\sqrt{3}}B\frac{(f_\pi-f_K)^2}{2f_K-f_\pi}\nonumber\\
&&+\frac{2\sqrt{2}}{\sqrt{3}}A\left( \frac{m_q}{f_\pi}+\frac{2m_s}{2f_K-f_\pi} \right) \label{me8v2}\\
m_{\eta_0\eta_8}^2&=&\frac{\sqrt{2}}{3}\lambda\left<{\sigma}_8\right>(\sqrt{2}\left<{\sigma}_0\right>-\frac{\left<{\sigma}_8\right>}{2})-2B\left<{\sigma}_8\right>,
\end{eqnarray}
where $m_{\sigma_0\sigma_8}^2$ and $m_{\eta_0\eta_8}^2$ are the mixing terms of $\sigma_0\sigma_8$ and $\eta_0\eta_8$, respectively.
The physical mass is defined so as to diagonalize the mass term.
For $\eta$ and $\eta^\prime$, we have
\begin{eqnarray}
m_{\eta}^2&=&\frac{1}{2}\left( m_{{\eta}_0}^2+m_{{\eta}_8}^2-\sqrt{(m_{{\eta}_0}^2-m_{{\eta}_8}^2)^2+4m_{\eta_0\eta_8}^4}) \right)\nonumber\\
\label{mepv}\\
m_{{\eta}^{\prime}}^2&=&\frac{1}{2}\left(m_{{\eta}_0}^2+m_{{\eta}_8}^2+\sqrt{(m_{{\eta}_0}^2-m_{{\eta}_8}^2)^2+4m_{\eta_0\eta_8}^4})
\right).\nonumber\\
\label{mev}
\end{eqnarray}

\section{\label{appb}The determination of parameters}
We determine the parameters in the linear sigma model from the physical values in vacuum.
The sigma condensates can be determined from the meson decay constants
through Eqs.(\ref{fpi}) and (\ref{fk}):
\begin{eqnarray}
\left<\sigma_0\right>&=&\frac{1}{\sqrt{6}}(f_\pi+2f_K),\\
\left<\sigma_8\right>&=&\frac{2}{\sqrt{3}}(f_\pi-f_K).
\end{eqnarray}

Once the sigma condensates are fixed, $Am_q$ and $Am_s$ can be determined by the $\pi$ and $K$ meson masses Eqs.(\ref{mpiv2}) and (\ref{mkv2}).
\begin{eqnarray}
Am_q&=&\frac{f_\pi}{2\sqrt{6}}m_\pi^2\\
A(m_q+m_s)&=&\frac{f_K}{\sqrt{6}}m_K^2
\end{eqnarray}
This fixes the ratio of $m_q$ and $m_s$.
In the linear sigma model, the quark masses appear always with the parameter $A$.
For independent determination of $m_q, m_s$ and $A$, we introduce an explicit value $m_q=5$MeV to fix $A$ and $m_s$.

Once noticing $m_{\eta_0}^2+m_{\eta_8}^2=m_{\eta^\prime}^2+m_\eta^2$ from Eqs.(\ref{mepv}) and (\ref{mev}), we can determine $B$ from $m_{\eta^\prime}^2+m_\eta^2$ with Eqs.(\ref{me0v2}) and (\ref{me8v2}),
\begin{eqnarray}
B&=&\frac{1}{\sqrt{6}} \frac{2f_K-f_\pi}{3f_\pi^3-8f_\pi f_K+8f_K^2}\nonumber\\
&\times&\left[(m_\eta^2+m_{\eta^\prime}^2)^2-2\sqrt{6}A\left( \frac{m_q}{f_\pi}+\frac{m_s}{2f_K-f_\pi} \right) \right]  
\end{eqnarray}
From Eqs.(\ref{mpiv4}) and (\ref{mkv4}), we can fix $\lambda$ from
\begin{eqnarray}
\lambda=\frac{m_K^2-m_\pi^2}{(f_K-f_\pi)(2f_K-f_\pi)}-\frac{2\sqrt{6}B}{2f_K-f_\pi}
\end{eqnarray}
Finally from the vacuum conditions Eqs.(\ref{vc3}) and (\ref{vc4}), we can fix $\mu^2$ and $\lambda^\prime$ as
\begin{eqnarray}
\mu^2 = a_1 \lambda +a_2 \lambda^\prime +a_3 B \label{mu2}
\end{eqnarray}
with
\begin{eqnarray}
a_1 &=& -\frac{m_8 (2\left<\sigma_0\right>^3 +6\left<\sigma_0\right> \left<\sigma_8\right>^2
 -\sqrt{2}\left<\sigma_8\right>^3)/6}{\left<\sigma_0\right>m_8 -\left<\sigma_8\right> m_0}\nonumber\\
&&+\frac{m_0 \left<\sigma_8\right>
 (2\left<\sigma_0\right>^2-\sqrt{2}\left<\sigma_0\right>
 \left<\sigma_8\right>
 +\left<\sigma_8\right>^2)/2}{\left<\sigma_0\right> m_8
 -\left<\sigma_8\right>m_0}\nonumber\\ \\
a_2 &=& -(\left<\sigma_0\right>^2+\left<\sigma_8\right>^2) \\
a_3 &=& \frac{2(m_8 (\left<\sigma_0\right>^2 -\frac{\left<\sigma_8\right>^2}{2}+m_0 \left<\sigma_8\right> (\left<\sigma_0\right>+\frac{\left<\sigma_8\right>}{\sqrt{2}})))}{\left<\sigma_0\right> m_8 -\left<\sigma_8\right>m_0}\nonumber\\
\end{eqnarray}
and
\begin{eqnarray}
 \lambda^\prime=\frac{m_\sigma^2-\lambda (a_1+\left<\sigma_0\right>^2+\left<\sigma_8\right>^2)-B(a_3-4\left<\sigma_0\right>)}{2\left<\sigma_0\right>^2}\nonumber\\
\end{eqnarray}

We show the input values to determine the parameters of the lagrangian and the determined parameters in TABLE \ref{input} and \ref{output}.
\begin{table*}
\caption{Input values}
\label{input}
\centering
\begin{tabular}{ccccccc}
\hline
$f_\pi$&$f_K$&$m_\pi$&$m_K$&$m_{\eta^\prime}^2+m_\eta^2$&$m_{\sigma_0}$&$m_q$\\
$[$MeV$]$&$[$MeV$]$&$[$MeV$]$&$[$MeV$]$&$[$MeV$^2 ]$&$[$MeV$]$&$[$MeV$]$\\
\hline \hline
92.2&110.4&135&495&550$^2$+958$^2$&700&5.0\\
\hline
\end{tabular}
\end{table*}

\begin{table*}
\caption{Determined quantities}
\label{output}
\centering
\begin{tabular}{cccccccccccccc}
\hline
$\left<\sigma_0\right>$&$\left<\sigma_8\right>$&$\mu^2$&$\lambda$&$\lambda^\prime$&$B$&$A$&$m_s$&$g$&$m_\eta$&$m_{\eta^\prime}$&$m_{\sigma_8}$&$(-\left< \bar{q}q \right>)^{1/3}$&$(-\left< \bar{s}s \right>)^{1/3}$\\
$[$MeV$]$&$[$MeV$]$&$[$MeV$^2]$&$[$ - $]$&$[$ - $]$&$[$MeV$]$&$[$MeV$^2]$&$[$MeV$]$&$[$ - $]$&$[$MeV$]$&$[$MeV$]$&$[$GeV$]$&$[$MeV$]$&$[$MeV$]$\\
\hline \hline
$128$&$-21.0$&$1.16\times 10^5$&$59.4$&$-2.4$&$984$&$6.86\times 10^4$&$156$&$7.67$&$535$&$959$&$1.23$&$249$&$279$\\
\hline
\end{tabular}
\end{table*}

\section{\label{appd}The calculation of the in-medium nucleon loop diagrams}
In this section, we show the explicit calculation of the nucleon one loop contribution to the sigma effective potential and the meson masses.
Here we assume the chiral limit.
We use the in-medium nucleon propagator defined as
\begin{eqnarray}
P_{\rm
 med}(p)&=&(\Slash{p}+m_N)\left\{\frac{i}{p^2-m_N^2+i\epsilon}\right.\nonumber\\
&&\left.-2\pi\delta(p^2-m_N^2)\theta(p_0)\theta(k_f-|\vec{p}|)\right\}
\end{eqnarray}
First, we evaluate the tadpole diagram for the $\sigma$ effective action denoted as $V_{MF}(\rho)$ in
Sec.\ref{med1}.
The effective potential for $\sigma_0$ coming with the nucleon-tadpole diagram $V_{MF}^0(\rho)$ is calculated as
\begin{equation}
-iV_{MF}^0(\rho)=-2\sigma_0g_{\sigma_0 NN}\int\frac{d^4p}{(2\pi)^4}{\rm tr}P_{\rm med}(p)
\end{equation}
with the $\sigma_0 N$ coupling
\begin{equation}
g_{\sigma_0 NN}=-\frac{ig}{\sqrt{3}}
\end{equation}
obtained from the lagrangian.
The factor 2 comes from the isospin degeneracy and the minus sign comes from the fermion loop.
Removing the pure vacuum contribution, which is divergent and should be renormalized into physical quantities, we have obtained as
\begin{equation}
V_{MF}^0(\rho)=\frac{g\rho}{\sqrt{3}}\sigma_0.\label{s0exch}
\end{equation}
Here we have used
\begin{eqnarray}
-\int\frac{d^4p}{(2\pi)^4}{\rm tr}P_{\rm med}(p)&=&\frac{4m_N}{(2\pi)^3}\int d^4p\frac{\delta(p_0-E_N)}{2E_N(\vec{p})}\theta(k_f-|\vec{p}|)\nonumber\\
&=&\frac{k_f^3}{3\pi^2}=\frac{\rho}{2},
\end{eqnarray}
where $\rho=\frac{2k_f^3}{3\pi^2}$ and $E_N(\vec{p})=\sqrt{|\vec{p}|^2+m_N^2}$.
Here we have approximated $E_N=m_N$.
In the same way, the effective potential for $\sigma_8$ is obtained as
\begin{eqnarray}
-iV_{MF}^8(\rho)&=&-2g_{\sigma_8 NN}\int\frac{d^4 p}{(2\pi)^4}{\rm tr}P_{\rm med}(p)\nonumber\\
&=&-i\frac{g\rho}{\sqrt{6}}\sigma_8\label{s8exch}
\end{eqnarray}
with the $\sigma_8 N$ coupling
\begin{equation}
g_{\sigma_8 N}=-\frac{ig}{\sqrt{6}}.
\end{equation}
Summing up Eqs.(\ref{s0exch}) and (\ref{s8exch}), we obtain Eq.(\ref{vmf}).

Next, we calculate the particle-hole contribution to the meson self-energy $\Sigma_{ph}(\rho)$.
The particle-hole contribution to the in-medium self-energy of mesons written as $\Sigma_{ph}(\rho)$ is
\begin{eqnarray}
-i\Sigma_{ph}(\rho) &=& -g^2C_i\int \frac{d^4p}{(2\pi)^4}{\rm tr}\{ {\gamma}_5P_{\rm med}(p+q)\nonumber\\
&&\times {\gamma}_5P_{\rm med}(p) \}
\end{eqnarray}
The coefficient $C_i$ is dependent on the channel;
$C_\pi=\frac{1}{2},C_{\eta_0}=\frac{1}{3},C_{\eta_8}=\frac{1}{6},C_{\eta_0\eta_8}=\frac{1}{3\sqrt{2}}$, which are obtained by the meson-nucleon couplings $g_{\pi NN}=g/\sqrt{2}$, $g_{\eta_0 NN}=g/\sqrt{3}$, $g_{\eta_8 NN}=g/\sqrt{6}$.
Denoting the part of the nucleon loop integral in $\Sigma_{ph}(\rho)$ as $\Pi(\rho)$ and removing the divergent vacuum contribution, we evaluate $\Pi(\rho)$ as follows;
\begin{eqnarray}
-i\Pi(\rho)&=&-\int \frac{d^4p}{(2\pi)^4}{\rm
 tr}[\gamma_5(\Slash{p}+\Slash{q}+m_N)\nonumber\\
&&\times \gamma_5(\Slash{p}+m_N)]\nonumber\\
&&\times\frac{i}{(p+q)^2-m_N^2+i\epsilon} \nonumber\\
& &\times (-2\pi )\delta \left( p^2-m_N^2
			 \right)\nonumber\\
&&\times \theta (p_0)\theta (k_f-|\vec{p}|) \nonumber\\
&=&-\frac{i}{(2\pi)^3}\int d^3\vec{p}\frac{4p \cdot q}{2p
 \cdot q+q^2}\frac{1}{2E_N(\vec{p})}\theta (k_f-|\vec{p}|) \nonumber\\
&=&-\frac{i}{(2\pi)^3} \int d^3 \vec{p} \frac{\theta (k_f-|\vec{p}|)}{m_N} \nonumber\\
&=&-\frac{i}{4m_N}\rho
\end{eqnarray}
From the first line to the second line, we used the Dirac equation and from the second line to the third line, we have taken the soft limit, $q^2=0$.
Here, we have omitted the contribution from the pure medium contribution, which contains the two step functions, because the contribution vanishes in the soft limit.
Multiplying the symmetry factor and the isospin degeneracy and adding the contribution from
another cross term of particle-hole diagram and the contribution from
crossed diagram, which gives the same contribution as the non-crossed
diagram in the soft limit, we obtain finally
\begin{eqnarray}
\Sigma_{ph}(\rho)=C_i\frac{g^2\rho}{m_N}
\end{eqnarray}

\section{\label{appe}Definition of scattering length}
Based on Ref.\cite{Ikeda2011}, the scattering length $a$ and effective range $r_e$ are given by the scattering amplitude $f(k)$ as
\begin{eqnarray}
f(k)&=&-\frac{M}{4\pi W}t(k) \label{scattamp}\\
a&=&\left. f(k)\right|_{k\rightarrow 0} \\
r_e&=&\left. \frac{d^2}{dk^2} \left( \frac{1}{f(k)} \right) \right|_{k\rightarrow 0}.
\end{eqnarray}
where $t(k)$ is the $T$-matrix defined in Eq.(\ref{lseq}).
The relation of the center-of-mass momentum $k$ and the total energy $W$ is
\begin{equation}
k=\frac{\sqrt{(W^2-(M+m)^2)(W^2-(M-m)^2)}}{2W}
\end{equation}
with the baryon mass $M$ and the meson mass $m$.


\begin{thebibliography}{20}
\bibitem{Weinberg1975}
S. Weinberg, Phys. Rev. {\bf D11} (1975) 3583.
\bibitem{Witten1979}
E. Witten, Nucl. Phys. {\bf B156} (1979) 269.
\bibitem{Veneziano1979}
G. Veneziano, Nucl Phys. {\bf B159} (1979) 213.
\bibitem{Bardeen1969}
W. A. Bardeen, Phys. Rev. {\bf 184} (1969) 1848.
\bibitem{Gross1981}
Gross, David J. and Pisarski, Robert D. and Yaffe, Laurence G., Rev. Mod. Phys. {\bf 53} (1981) 43.
\bibitem{Kapsta1996}
J. Kapsta, D. Karzeev, L. McLerran, Phys. Rev. {\bf D53} (1996) 5028.
\bibitem{Suzuki2004}
K. Suzuki, {\it et al}, Phys. Rev. Lett. {\bf 92} (2004) 72302.
\bibitem{Jido2012}
D. Jido, H. Nagahiro, S. Hirenzaki, Phys. Rev. {\bf C85} (2012) 032201(R).
\bibitem{Pisarski1984}
R.D. Pisarski, F. Wilczek, Phys. Rev. {\bf D29} (1984) 338.
\bibitem{Bernard1987}
V. Bernard, U.G. Meissner, Phys. Rev. {\bf D38} (1988) 1551.
\bibitem{Hatsuda1994}
T. Hatsuda, T. Kunihiro, Phys. Rep. {\bf 247} (1994) 221.
\bibitem{Tsushima2000}
K. Tsushima, Nucl. Phys. {\bf A670} (2000) 198.
\bibitem{Saito2007}
K. Saito, K. Tsushima, A.W. Thomas, Prog. Part. Nucl. Phys. {\bf 58}
	(2007) 1.
\bibitem{Costa2003}
P. Costa, M.C. Ruivo, Y.L. Kalinovsky, Phys. Lett. {\bf B569} (2003) 171.
\bibitem{Bass2006}
S.D. Bass, A.W. Thomas, Phys. Lett. {\bf B634} (2006) 368.
\bibitem{Nagahiro2005}
H. Nagahiro, S. Hirenzaki, Phys. Rev. Lett. {\bf 94} (2005) 232503
\bibitem{Nagahiro2012}
H. Nagahiro, S. Hirenzaki, E. Oset, A. Ramos, Phys. Lett. {\bf B709} (2012) 87.
\bibitem{Nagahiro2006}
H. Nagahiro, M. Takizawa, S. Hirenzaki, Phys. Rev. {\bf C74} (2006) 045203.
\bibitem{Oset2011}
E. Oset, A. Ramos, Phys. Lett. {\bf B704} (2011) 334.
\bibitem{Benic2011}
S. Beni\ifmmode \acute{c}\else \'{c}\fi{}, D. Horvati\ifmmode
	\acute{c}\else \'{c}\fi{}, D. Kekez, D. Klabu\ifmmode
	\check{c}\else \v{c}\fi{}ar, Phys. Rev. {\bf D84} (2011) 016006.
\bibitem{Kwon2012}
Y. Kwon, S.H. Lee, K. Morita, G. Wolf, Phys. Rev. {\bf D86} (2012) 034014.
\bibitem{Lee2013}
S.H. Lee, S. Cho, Int. J. Mod. Phys. {\bf E22} (2013) 1330008.
\bibitem{Moskal2000a}
P. Moskal, {\it et al}. Phys. Lett. {\bf B474} (2000) 416.
\bibitem{Moskal2000b}
P. Moskal, {\it et al}. Phys. Lett. {\bf B484} (2000) 356.
\bibitem{Nanova2012}
M. Nanova, {\it et al}. Phys. Lett. {\bf B710} (2012) 600.
\bibitem{Csorgo2010}
Cs\"org\ifmmode \mbox{\H{o}}\else \H{o}\fi{}, T. and V\'ertesi, R. and Sziklai, J., Phys. Rev. Lett. {\bf 105} (2010) 182301.
\bibitem{Cohen1996}
T.D. Cohen, Phys. Rev. {\bf D54} (1996) 1867.
\bibitem{Lee1996}
S.H. Lee, T. Hatsuda, Phys. Rev. {\bf D54} (1996) 1871.
\bibitem{Jido2012a}
D. Jido, {\it et al}, Nucl. Phys. {\bf A914} (2013) 354.
\bibitem{Shuryak1982}
E.V. Shuryak, Nucl. Phys.{\bf  B203} (1982) 140.
\bibitem{Drukarev1991}
E.G. Drukarev, E.M. Levin, Prog. Part. Nucl. Phys. {\bf 27} (1991) 77.
\bibitem{Gell-Mann1960}
M. Gell-Mann, M. Levy, IL Nuovo Cim. {\bf 16} (1960) 705.
\bibitem{Schechter1971}
J. Schechter, Y. Ueda, Phys. Rev. {\bf D3} (1971) 168.
\bibitem{Kawarabayashi1980}
K. Kawarabayashi, N. Ohta, Nucl. Phys. {\bf B175} (1980) 477.
\bibitem{Lenaghan2000}
Lenaghan,J.T., Rischke,D.H., Schaffner-Bielich,J., Phys. Rev. {\bf D62} (2000) 085008.
\bibitem{Kobayashi1970}
M. Kobayashi, T. Maskawa, Prog. Theor. Phys. {\bf 44} (1970) 1422.
\bibitem{'tHooft1976}
G. 't Hooft, Phys. Rev. {\bf D14} (1976) 3432.
\bibitem{Christos1987}
G.A. Christos, Phys. Rev. {\bf D35} (1987) 330.
\bibitem{Lee1974}
T.D. Lee, G.C. Wick, Phys. Rev. {\bf D9} (1974) 2291.
\bibitem{DeTar1989}
C. DeTar, T. Kunihiro, Phys. Rev. {\bf D39}(1989) 2805.
\bibitem{Jido1998av}
D. Jido, Y. Nemoto, M. Oka, A. Hosaka, Nucl. Phys. {\bf A671} (2000) 471.
\bibitem{Kim1998upa}
H. -c. Kim, D.Jido, M.Oka, Nucl. Phys. {\bf A640} (1998) 77.
\bibitem{Sasaki2011}
Sasaki,C., Lee,H.K., Paeng,W.G., Rho,M., Phys. Rev. {\bf D84} (2011) 034011.
\bibitem{Oller1997}
J.A. Oller, E. Oset, Nucl. Phys. {\bf A620} (1997) 438.
\bibitem{Hyodo2008}
T. Hyodo, D. Jido, A. Hosaka, Phys. Rev. {\bf C78} (2008) 025203.
\bibitem{Ikeda2011}
Y. Ikeda, T. Hyodo, D. Jido, H. Kamano, T. Sato, K. Yazaki,
	Prog. Theor. Phys. {\bf 125} (2011) 1205.
\end{thebibliography}
\end{document}